\documentclass[symmetry,review,accept,pdftex,moreauthors]{Definitions/mdpi} 

\firstpage{1} 
\pubvolume{18}
\issuenum{1}
\articlenumber{200}
\pubyear{2026}
\copyrightyear{2026}
\externaleditor{Stefano Profumo} 
\datereceived{18 December 2025} 
\daterevised{15 January 2026} 
\dateaccepted{19 January 2026} 
\datepublished{21 January 2026}

\Title{A Review of Hyperon Physics at BESIII Experiment}

\Author{Ruoyu 
 Zhang $^{1,2}$\orcidA{} and Xiongfei Wang $^{1,2,3,}$*\orcidB{}}
 
\AuthorNames{Ruoyu Zhang and Xiongfei Wang}

\address{%
$^{1}$ \quad School of Physical Science and Technology, Lanzhou University, Lanzhou 730000, China;  {zhangry18@lzu.edu.cn} 
\\
$^{2}$ \quad Lanzhou {Center} for Theoretical Physics, Key Laboratory of Theoretical Physics of Gansu Province, Key~Laboratory of Quantum Theory and Applications of MoE, Gansu Provincial Research Center for Basic Disciplines of Quantum Physics, Lanzhou University, Lanzhou 730000, China\\

$^{3}$ \quad Frontiers Science Center for Rare Isotopes, Lanzhou University, Lanzhou 730000, China
}
\corres{Correspondence: wangxiongfei@lzu.edu.cn}

\abstract{The BESIII Collaboration has collected large data samples from $e^+e^-$ collisions at center-of-mass energies ranging from 1.84 to 4.95 GeV, which  include the world’s largest charmonium sample, consisting of 10 billion $J/\psi$ and 3 billion $\psi(3686)$ events. 
These high-statistics datasets enable BESIII to carry out a wide range of studies in hyperon physics.
In this article, we review the major achievements of the BESIII Collaboration in this field, which can be broadly categorized into four areas: hyperon polarization and $CP$ violation, rare hyperon decays, hyperon pair production, and hyperon–nucleon interactions.}


\begin{document}
\section{Introduction}
The Big Bang implies the production of equal amounts of matter and antimatter, which should have led to their complete annihilation in the early Universe~\cite{Dine:2003ax}. However, observations indicate that the present Universe is overwhelmingly composed of matter, with little evidence of antimatter. This discrepancy has puzzled the scientific community for more than half a century and remains a subject of intensive research. One of the long-standing explanations is baryogenesis~\cite{Sakharov:1967dj}, in which the matter abundance is generated dynamically. 
A crucial requirement for its realization is the violation of charge-conjugation and parity ($CP$) symmetry.
Therefore, exploring $CP$ violation has become one of the key objectives of modern collider experiments.
To date, the clear existence of $CP$ violation has been established in weak interactions involving kaon, charm, and beauty meson decays~\mbox{\cite{Christenson:1964fg,BaBar:2001pki,Belle:2001zzw,LHCb:2019hro}}. There is also experimental evidence for $CP$ violation in neutrino oscillations~\cite{T2K:2019bcf}.
The observed $CP$ violation in meson decays is well-interpreted by the Cabibbo--Kobayashi--Maskawa (CKM) mechanism~\cite{Cabibbo:1963yz,Kobayashi:1973fv}.
However, the matter–antimatter asymmetry generated by the CKM mechanism is far too small to account for astronomical observations, posing a serious challenge to the Standard Model (SM) and indicating the possible existence of additional sources of $CP$ violation~\cite{Dine:2003ax}. This has driven the expansion of $CP$ violation research into broader domains.
Very recently, LHCb reported the first observation of $CP$ violation in $\Lambda^0_b$ baryons~\cite{LHCb:2025ray}, opening a new avenue in the search for physics beyond the SM. 
The development of angular distribution analysis based on the helicity formalism has provided a new approach to exploring $CP$ violation in hyperons. 
The data samples of 10 billion $J/\psi$ and 3 billion $\psi(3686)$ events collected by BESIII in $e^+e^-$ collisions at center-of-mass (c.m.) energies ($\sqrt{s}$) of 3.097 and 3.686 GeV, respectively, are well-suited for hyperon $CP$ tests, and both states exhibit sizable branching fractions to hyperon–antihyperon pairs.
Moreover, $e^+e^-$ annihilation offers an experimental environment with a relatively low background and enables the production of quantum-entangled hyperon pairs.
These features create highly favorable conditions for studying hyperon $CP$ violation at BESIII~\cite{Wang:2023trb}.
Furthermore, angular distribution analysis not only enables the exploration of $CP$ violation but also provides a novel approach for accurately extracting hyperon form factors. Since the first experimental determination of the proton form factors in the 1950s~\cite{Punjabi:2015bba}, form factors have remained a fundamental property of baryons and have been extensively studied to this day. The angular distribution analysis allows the extraction of the modulus ratio of the form factors, thereby extending the study of form factors beyond the relatively stable proton and neutron and providing valuable insights into the internal structure of hyperons. In this respect, BESIII data collected at three energy regions, such as near the hyperon–antihyperon production threshold (2.396–2.900 GeV), around the $\psi(3686)$ resonance (3.68–3.71 GeV), and at the $\psi(3770)$ resonance (3.773 GeV), are used in the analyses.

Another way to probe physics beyond the SM is through searches for rare decays, including in the hyperon sector. Within and beyond the SM, many theoretical models predict baryon-number-violating (BNV) decays. Experimentally, extensive searches have been carried out, but no evidence for such processes has been observed so far. Among the theories, some predict BNV decays that obey $\Delta(B-L)=0$, where $\Delta(B-L)$ denotes the change in the difference between baryon and lepton numbers. 
As a hyperon factory, BESIII provides an opportunity to explore BNV decays in hyperons, while the presence of an $s$ quark coupled to a $u$ or $d$ quark in hyperons allows the study of possible interference among multiple amplitudes in BNV transitions, an effect that has not been taken into account in existing theoretical calculations due to the ample parameter space.
On the other hand, hyperon semileptonic decays provide valuable insight into the interplay between strong and weak interactions. The $\Delta S = \Delta Q$ rule is a fundamental assumption arising from the SU(3) symmetry of the weak hadronic current in the Cabibbo theory of weak interactions, introduced to explain the experimental absence of certain hyperon decay modes. Here, $\Delta S$ and $\Delta Q$ denote the changes in strangeness and charge between the initial and final state hadrons, respectively. Searching for hyperon semileptonic decays that violate the $\Delta S = \Delta Q$ rule would, therefore, signal the presence of weak currents associated with higher SU(3) multiplets.
To ensure sufficient statistics, the full data sample of $J/\psi$ data collected by BESIII is used in these studies.

At BESIII, substantial data samples have also been collected from the hyperon pair production threshold up to the energy region of the $XYZ$ states~\cite{Brambilla:2019esw}, ranging from 1.84 to 4.95 GeV, providing valuable opportunities to study hyperon pair production in $e^+e^-$ annihilation, to further explore the nature of charmonium(-like) states, and to test the nonperturbative theory of quantum chromodynamics.
Experimental observations of the vector charmonium(-like) states exceed the predictions of potential models. This difference provide a significant opportunity to probe exotic configurations of quarks and gluons~\cite{Wang:2025dur}.
A variety of theoretical interpretations, including hybrid, multiquark, and molecular configurations~\cite{Farrar, Briceno, Chen:2016qju, Close:2005iz, Wang:2019mhs, Qian:2021neg, BESIII:2024ths}, have been proposed to explain charmonium(-like) states. Nevertheless, no definitive conclusion has been reached, and the underlying nature of these states remains unresolved. This situation highlights our incomplete understanding of strong interaction dynamics in the nonperturbative regime. Further progress therefore relies on more precise experimental measurements. In this context, investigations of hyperon decays of charmonium(-like) states are particularly attractive, owing to their simple final state topologies and comparatively well-understood decay mechanisms. Additionally, the discovery of charmonium(-like) states in $e^+e^-$ annihilation into charmonium and light hadrons highlights the importance of studying hyperon final states, where information is still limited~\cite{Wang:2024gtk}.

In addition, scattering experiments play a central role in probing the fundamental forces and internal structure of matter. Over the years, various types of particle beams have been produced and applied in scattering experiments, driving numerous scientific breakthroughs. However, for other neutral particles that are experimentally difficult to control, as well as long-lived hyperons and their antiparticles, experimental measurements remain scarce, even though they possess considerable physical potential.
As a result, precise modeling of hyperon–nucleon and hyperon–hyperon interactions remains difficult, despite the availability of strong constraints and well-established descriptions for nucleon–nucleon interactions~\cite{Vidana:2018bdi,Hiyama:2018lgs}. The properties of hyperons in dense matter have therefore drawn significant attention, owing to their close relevance to hypernuclei and to the role of hyperons in neutron stars~\cite{Tolos:2020aln}.
In neutron stars, the possible presence of hyperons decreases the Fermi pressure of the system in the equation of state (EoS), thereby reducing the maximum mass that a neutron star can sustain. EoS calculations that include hyperons often predict a maximum mass lower than the observed values—a discrepancy known as the ``hyperon puzzle''~\cite{Lattimer:2000nx,Lonardoni:2014bwa,LIGOScientific:2018cki}, which further motivates the exploration of hyperon–nucleon interactions.
Based on the abundant hyperon data produced in $J/\psi$ decays, BESIII has pioneered a series of hyperon–nucleon scattering studies using the nucleons in the beam-pipe material as~targets.

In this article, we review recent results from the BESIII experiment related to hyperon polarization and $CP$ violation, rare hyperon decays, hyperon production cross sections, and studies of hyperon–nucleon scattering.

\section{The BESIII Detector}
The BESIII detector~\cite{Ablikim:2009aa} has recorded data from symmetric electron–positron collisions provided by the BEPCII storage ring~\cite{Yu:IPAC2016-TUYA01}, covering c.m. energies from 1.84 to 4.95 GeV. At $\sqrt{s} = 3.773\;\text{GeV}$, it achieved a peak luminosity of $1.1 \times 10^{33}\;\text{cm}^{-2}\text{s}^{-1}$. BESIII has accumulated a large data sample in this energy region~\cite{Ablikim:2019hff,EcmsMea,EventFilter}.

The cylindrical core of the Beijing Spectrometer III (BESIII) detector covers 93\% of the solid angle and consists of a helium-based multi-wire drift chamber~(MDC), a time-of-flight (TOF) detector, and a CsI(Tl)  electromagnetic calorimeter~(EMC). These detecting components are housed within a superconducting solenoid magnet that provides a magnetic field of 1.0 Tesla (0.9 Tesla in 2012). The solenoid is supported by an octagonal yoke, which incorporates resistive plate chambers for muon identification interleaved with steel plates.
The detection efficiency for charged particles and photons within a $4\pi$ solid angle is 93\%. At a momentum of 1 GeV$/c$, the momentum resolution for charged particles is 0.5\%, and for electrons from Bhabha scattering, the ${\rm d}E/{\rm d}x$ resolution is 6\%. The EMC achieves an energy resolution of 2.5\% (5\%) for 1 GeV photons in the barrel (end-cap) region. The plastic scintillator TOF detector has a time resolution of 68 ps in the barrel, while it was originally 110 ps in the end-cap region. The end-cap system was upgraded in 2015 using multi-gap resistive plate chamber technology, improving the time resolution to 60 ps and effectively enhancing data quality~\cite{etof1,etof2,etof3}.

\section{Hyperon Polarization and \textit{CP} Violation}
\subsection{Analysis Method}
\begin{figure}[H]
 \includegraphics[width=1.0\textwidth]{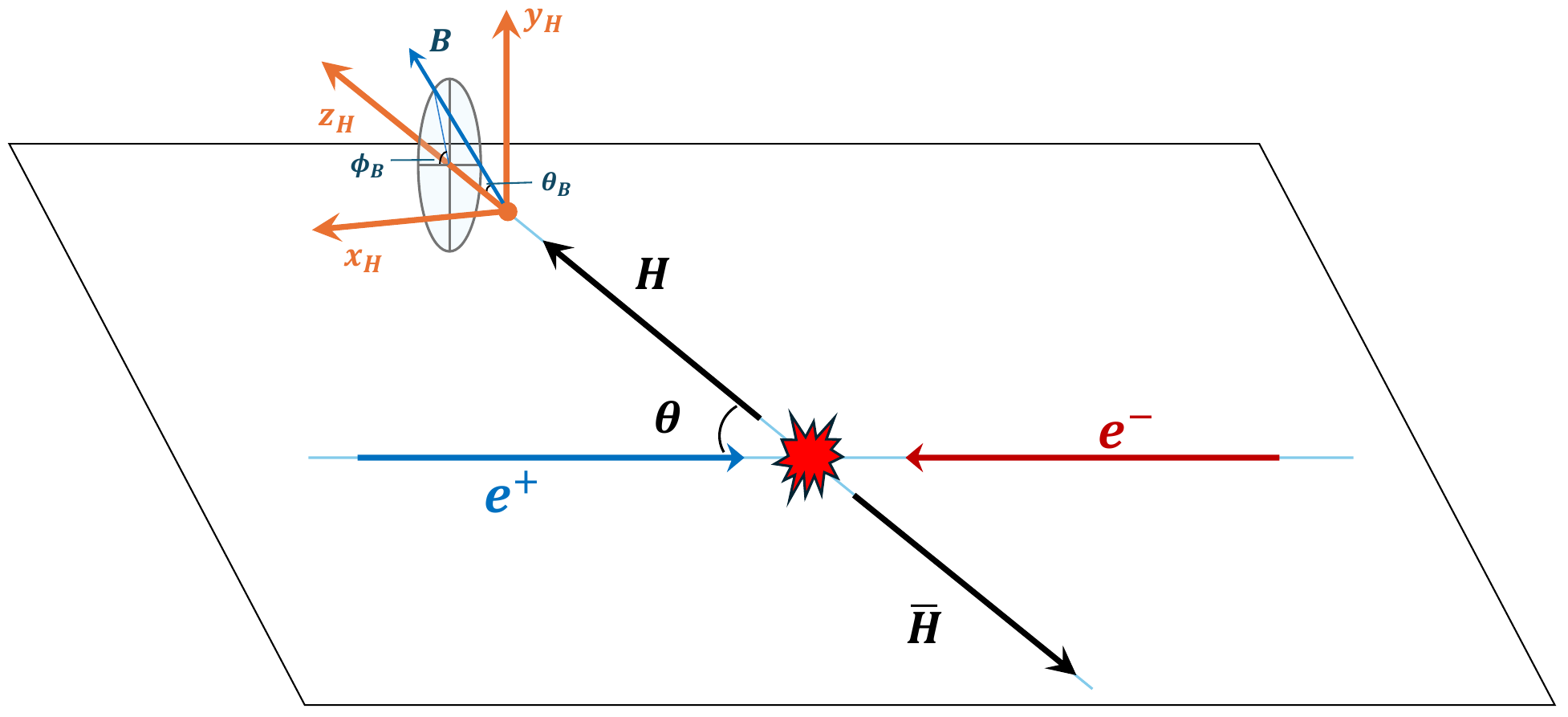}
\caption{\label{fig:cs} Orientation axes of the hyperon $H$ and antihyperon $\bar{H}$ helicity frames. $B$ represents the baryon produced by $H$ decay.}
\end{figure}

The $CP$ asymmetry is usually probed by comparing the decay patterns of a particle and its antiparticle. At BESIII, it is studied through angular distribution analysis within the helicity formalism, using decay parameters as observables.
The interaction vertex of quantum-entangled spin-1/2 hyperon–antihyperon pairs produced via $e^+e^-$ annihilation can be described by two electromagnetic form factors (EMFFs), $G_M$ and $G_E$~\cite{Faldt:2013gka,Faldt:2017kgy}:
\begin{equation}
\Gamma_{\mu}(p_1,p_2) = -ie_g \left[ G_M \gamma^{\mu} - \frac{2M}{Q^2} (G_M - G_E) Q_{\mu} \right],
\label{vtx}
\end{equation}
where $M$ is the hyperon mass, $p_1$ and $p_2$ are the hyperon and antihyperon momenta, respectively, $Q = p_1 - p_2$, and $e_g$ is the coupling strength. Based on this, the angular distribution can be characterized by two real parameters: one corresponding to the ratio of the absolute values of the EMFFs, and the other representing their relative phase:
\begin{equation}
\begin{aligned}
\alpha_{\psi} &=
\frac{s|G_M|^2 - 4M^2|G_E|^2}
     {s|G_M|^2 + 4M^2|G_E|^2}, \\
\frac{G_E}{G_M} &=
e^{i\Delta\Phi}\left|\frac{G_E}{G_M}\right|.
\end{aligned}
\label{eq:eff}
\end{equation}
The production via charmonium states follows the same formalism, with the corresponding form factors, $G_E^{\psi}$ and $G_M^{\psi}$, referred to as psionic form factors~\cite{Faldt:2017kgy}.
Based on the optical theorem, the form factors in the space-like region are real at the lowest order as a consequence of the Hermiticity of the electromagnetic Hamiltonian, whereas in the time-like region they become complex, giving rise to transverse polarization of the produced hyperons~\cite{Dubnickova:1992ii,Tomasi-Gustafsson:2005svz,Denig:2012by}.
In general, by defining the coordinate system as shown in Figure~\ref{fig:cs}, the polarization terms can be expressed as follows~\cite{Perotti:2018wxm,BESIII:2019qjb}:
\begin{equation}
P_y(\cos\theta) = \frac{\sqrt{1 - \alpha_{\psi}^2 \sin(\Delta\Phi) \cos\theta \sin\theta}}{1 + \alpha_{\psi} \cos^2\theta},
\end{equation}
while a projection variable $\mu(\cos\theta)$ (or $M(\cos\theta)$) is often defined experimentally to illustrate the model’s description of data and reveal the polarization signal:
\begin{equation}
\mu(\cos\theta) = \frac{m}{N} \sum^{N_k}{i=1} (n_{1,y}^{(i)} - n_{2,y}^{(i)}),
\end{equation}
where $m$ represents the bin number in $\cos\theta$, $N$ is the total number of events in the data sample, $N_k$ is the number of events in the $k$-th $\cos\theta$ bin, and {$\hat{\mathbf{n}}_1$ ($\hat{\mathbf{n}}_2$)}  
 denotes the unit vector along the direction of the daughter (antidaughter) particle in the rest frame of the hyperon (antihyperon).\vspace{-6pt}

Subsequently, the decay amplitude of a spin-$1/2$ hyperon can be wirtten as a combination of $S$-wave and $P$-wave contributions,
\begin{align}
S &= |S| e^{i(\delta_S + \xi_S)}, \\ \notag
P &= |P| e^{i(\delta_P + \xi_P)},
\end{align}
where $\delta_{S(P)}$ and $\xi_{S(P)}$ denote the strong and weak phases, respectively.
These amplitudes are associated with $CP$ conservation and violation, and are characterized by the Lee--Yang parameters $\alpha_H$, $\beta_H$, and $\gamma_H$~\cite{Lee:1957qs}, which satisfy the normalization condition $\alpha_H^2\nobreak+\nobreak\beta_H^2\nobreak+\nobreak\gamma_H^2\nobreak=\nobreak1$.
They are further expressed by defining a parameter $\phi_H\nobreak=\nobreak\tan^{-1}(\beta_H/\gamma_H)$. If $CP$ is conserved, the above hyperon decay parameters should be equal in magnitude but opposite in sign to those of their corresponding antiparticles. This allows us to define observables for $CP$ violation:
\begin{align}
    A_{CP} &= \frac{\alpha_H + \alpha_{\bar H}}{\alpha_H - \alpha_{\bar H}} = \frac{\beta_H + \beta_{\bar H}}{\beta_H - \beta_{\bar H}} = \frac{\gamma_H + \gamma_{\bar H}}{\gamma_H - \gamma_{\bar H}}, \\ \notag
    \Delta\phi_{CP} &= \frac{\phi_H + \phi_{\bar H}}{2}.
\end{align}
In addition, based on the strong ($\delta_P-\delta_S$) and weak ($\xi_P-\xi_S$) phase differences between the $S$-wave and $P$-wave, the $CP$ violation observables can be further expressed as~\cite{Donoghue:1985ww,Donoghue:1986hh}
\begin{equation}
    A_{CP} \approx -\tan(\delta_P-\delta_S)\tan(\xi_P-\xi_S).
\end{equation}

\subsection{Charmonium Decay}
\begin{figure}[H]

\begin{adjustwidth}{-\extralength}{0cm}
\centering 
{\includegraphics[width=1.0\textwidth]{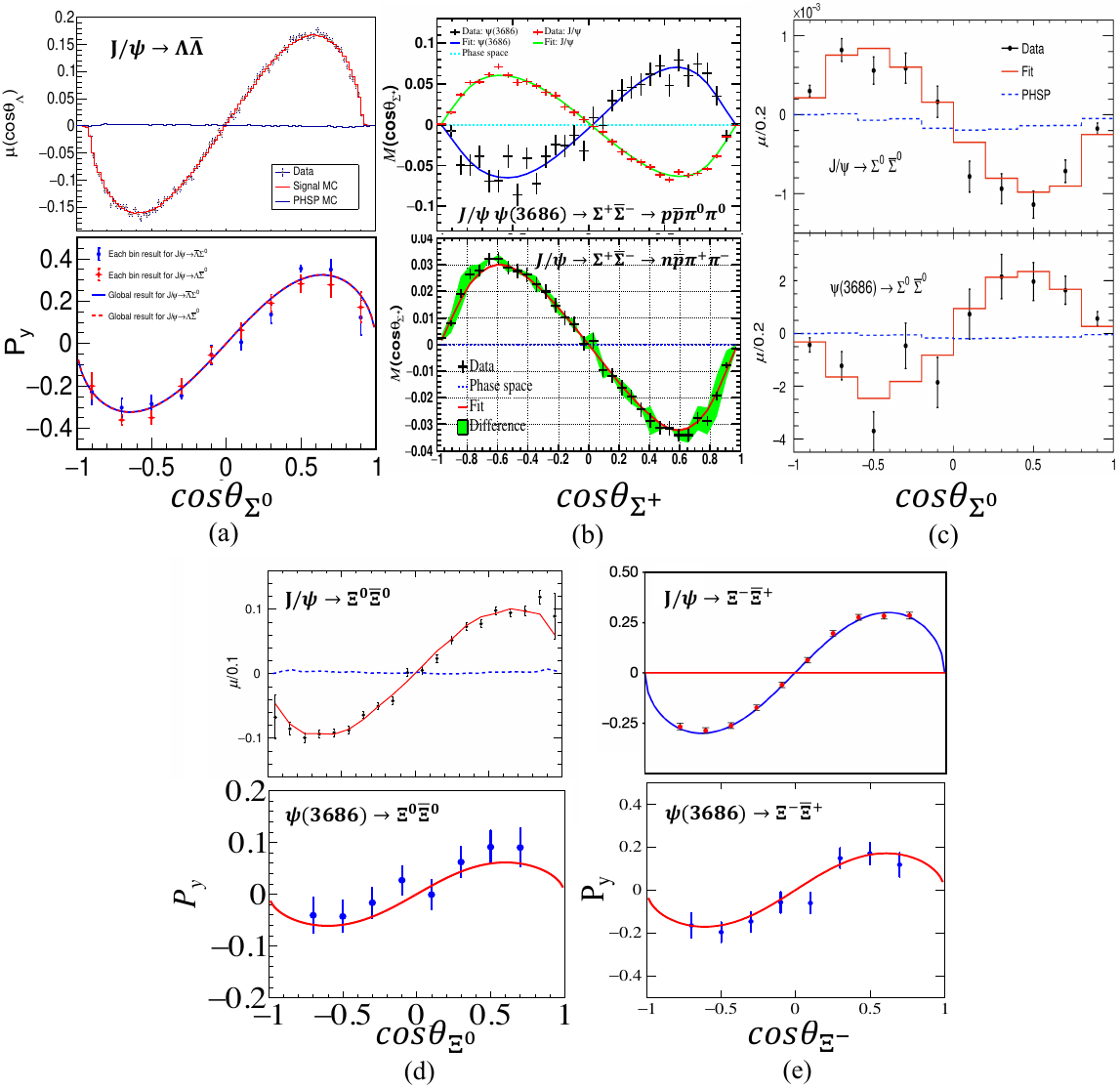}}
\end{adjustwidth}
\caption{\label{fig:polarization} 
{Distributions} of the polarization observables $\mu$ for $J/\psi \to \Lambda\bar{\Lambda}$~\cite{BESIII:2022qax} (\textbf{a}), $J/\psi/\psi(3686) \to \Sigma^0\bar{\Sigma}^0$~\cite{BESIII:2024nif} (\textbf{c}), and $J/\psi \to \Xi^0\bar{\Xi}^0$~{\cite{BESIII:2023drj}} (\textbf{d}); $M$ for $J/\psi/\psi(3686) \to \Sigma^+\bar{\Sigma}^-$~\cite{BESIII:2025jxt,BESIII:2023sgt} (\textbf{b}); and $P_y$ for $J/\psi \to \Lambda\Sigma^0 + \text{c.c.}$~\cite{BESIII:2023cvk} (\textbf{a}), $\psi(3686) \to \Xi^0\bar{\Xi}^0$~\cite{BESIII:2025dke} (\textbf{d}), and $J/\psi/\psi(3686) \to \Xi^-\bar{\Xi}^+$~\cite{BESIII:2021ypr,BESIII:2022lsz} (\textbf{e}), as functions of $\cos\theta$ for each process.
The dots with error bars represent the experimental data, the colored solid curves denote the global fit results, the dashed histograms correspond to the phase space (PHSP) Monte Carlo (MC) samples, and the red solid histograms show the PHSP MC weighted according to the global fit.
}

\end{figure}

BESIII has recently reported a series of measurements on hyperon spin polarization, aiming to explore possible indications of $CP$ violation. These measurements are based on the world's largest $J/\psi$ and $\psi(3686)$ data samples. Due to differences in data-taking periods, some analyses were performed using only partial datasets. 
The most precise measurement of the $\Lambda$ decay parameters was performed using the process $J/\psi \to \Lambda \bar{\Lambda}$~\cite{BESIII:2022qax}, as shown in Figure~\ref{fig:polarization}a. A clear polarization signal was observed, and the measurements of the decay parameters $\alpha^{\Lambda}_-$ for $\Lambda \to p\pi^-$ and $\alpha^{\bar\Lambda}_+$ for $\bar{\Lambda} \to \bar{p}\pi^+$ achieved the highest precision to date, with results of $\alpha^{\Lambda}_- = 0.7519\pm0.0036\pm0.0024$ and $\alpha^{\bar\Lambda}_+ = -0.7559\pm0.0036\pm0.0030$, respectively. The $CP$ violation observable for the $\Lambda$ decay was determined to be $A^{-}_{CP} = 0.0025 \pm 0.0046 \pm 0.0012$, which is consistent with zero, indicating no evidence of $CP$ violation. Meanwhile, the angular distribution parameters $\alpha_{J/\psi}$ and $\Delta\Phi$ are also determined to be $\alpha_{J/\psi} = 0.4748\pm0.0022\pm0.0031$ and $\Delta\Phi = 0.7521\pm0.0042\pm0.0066$~rad. 

The isospin-violating process $J/\psi \to \Lambda \bar{\Sigma}^0$, together with its charge-conjugation ($\text{c.c.}$), can also be studied through joint angular distribution analysis~\cite{BESIII:2023cvk}. The fit results of $P_y$ in each $\cos\theta_{\Sigma^0/\bar\Sigma^0}$ bin are shown in Figure~\ref{fig:polarization}a. Unlike hyperon–antihyperon pair production, non-charge-conjugated final states $\Lambda$ and $\bar{\Sigma}^0$ enable exploration of direct $CP$ violation by comparing the polarizations extracted from both decay modes.
The measurement extracted $\alpha_{J/\psi} = 0.418 \pm 0.028 \pm 0.014$, $\Delta\Phi_{\bar\Lambda\Sigma^0} = 1.011 \pm 0.094 \pm 0.010$~rad, and $\Delta\Phi_{\Lambda\bar{\Sigma}^{0}} = 2.128 \pm 0.094 \pm 0.010$~rad. Furthermore, the modulus ratio of the form factors was determined to be $R = 0.860 \pm 0.029 \pm 0.015$, and the $CP$ violation observable, $\Delta\Phi_{CP} = |\pi - (\Delta\Phi_{\bar{\Lambda}\Sigma^{0}} + \Delta\Phi_{\Lambda\bar{\Sigma}^{0}})| = 0.003 \pm 0.133 \pm 0.014$, is consistent with zero, indicating no significant evidence for direct $CP$ violation.

For the $\Sigma^+\bar{\Sigma}^-$ hyperon pair, based on both the $J/\psi$ and $\psi(3686)$ data samples, BESIII first reconstructed the decays via the dominant channels $\Sigma^+ \to p\pi^0$ and $\bar{\Sigma}^- \to \bar{p}\pi^0$, and measured their decay parameters~\cite{BESIII:2020fqg,BESIII:2025jxt}. 
Subsequently, by reconstructing one side of the decay through the neutron channel, the decay parameters of $\Sigma^+ \to n\pi^+$ and $\bar{\Sigma}^- \to \bar{n}\pi^-$ were measured~\cite{BESIII:2023sgt}, representing the first test of $CP$ symmetry in hyperon decays involving a neutron final state. The ratio of the decay asymmetry parameters for the two isospin decay modes serves as a sensitive probe for determining the contribution of $\Delta I = 3/2$ transitions and for investigating the $\Delta I = 1/2$ rule~\cite{Olsen:1970vb}. As observables, the ratios of the decay asymmetry parameters, $\alpha^{\Sigma^+}_+/\alpha^{\Sigma^+}_0$ and $\alpha^{\bar\Sigma^-}_-/\alpha^{\bar\Sigma^-}_0$, are found to be smaller than unity by more than $5\sigma$, indicating the presence of a $\Delta I = 3/2$ transition.

The decay parameters of the radiative decay $\Sigma^0 \to \gamma\Lambda$ offer an opportunity to explore parity and strong $CP$ violation beyond the SM~\cite{Nair:2018mwa}. In this case, the $CP$ violation observable is defined as $A_{CP}^{\Sigma^0} = \alpha^{\Sigma^0} + \alpha^{\bar{\Sigma}^0}$. Based on both $J/\psi$ and $\psi(3686)$ data samples, BESIII has also carried out a related study~\cite{BESIII:2024nif}.
The aforementioned measurements of the $\Sigma$ hyperons are summarized in Table~\ref{tab:sigma}, with results consistent with $CP$ conservation. The observed polarization signals are shown in Figure~\ref{fig:polarization}b,c.

\begin{table}[H]	
\footnotesize
\caption{{Fitting} 
 results of each decay parameter for $J/\psi/\psi(3686)\to\Sigma^+\bar\Sigma^-$ and $J/\psi/\psi(3686)\to\Sigma^0\bar\Sigma^0$. The first uncertainty is statistical and the second is systematic. The parameters $\alpha^{\Sigma^+}_0$, $\alpha^{\Sigma^+}_+$, and $\alpha^{\Sigma^0}$ denote the decay asymmetry parameters for the processes $\Sigma^+ \to p\pi^0$, $\Sigma^+ \to n\pi^+$, and $\Sigma^0 \to \gamma\Lambda$, respectively. $A_{CP}^{\Sigma}$ represents the $CP$ observable, determined from the corresponding decay~parameters.}

\begin{adjustwidth}{-\extralength}{0cm}
        \centering
   \setlength{\tabcolsep}{0.83mm}{\begin{tabular}{l c c c}
  \toprule
        \textbf{Parameter} 	 &~~\boldmath{$J/\psi/\psi(3686) \to \Sigma^+\bar\Sigma^- \to p\pi^0\bar{p}\pi^0$}~\textbf{\cite{BESIII:2025jxt}}~~     &~~\boldmath{$J/\psi \to \Sigma^+\bar\Sigma^- \to n\pi^+(p\pi^0)\bar{p}\pi^0(\bar{n}\pi^-)$}~\textbf{\cite{BESIII:2023sgt}}~~       &~~\boldmath{$J/\psi/\psi(3686) \to \Sigma^0\bar\Sigma^0 \to \gamma\Lambda\gamma\bar\Lambda$}~\textbf{\cite{BESIII:2024nif}}\\
    \midrule
        $\alpha_{J/\psi}$      	                            &$-0.5047\pm0.0018\pm0.0010$                      &$-0.5156\pm0.0030\pm0.0061$                    &$-0.4133\pm0.0035\pm0.0077$                        \\
        $\Delta\Phi_{J/\psi}$(rad)                          &$-0.2744\pm0.0033\pm0.0010$                      &$-0.2772\pm0.0044\pm0.0041$                    &$-0.0828\pm0.0068\pm0.0033$                        \\
        $\alpha_{\psi(3686)}$                               &$0.7133\pm0.0094\pm0.0065$       &$-$                                            &$0.814\pm0.028\pm0.028$            \\
        $\Delta\Phi_{\psi(3686)}$(rad)                      &$0.427\pm0.022\pm0.003$          &$-$                                            &$0.512\pm0.085\pm0.034$            \\
                            
        $\alpha^{\Sigma^+}_0$                               &$-0.975\pm0.011\pm0.002$                         &$-$                                            &$-$                                                \\
        $\alpha^{\bar\Sigma^-}_0$                           &$0.999\pm0.011\pm0.004$          &$-$                                            &$-$                                                \\
                    
        $\alpha^{\Sigma^0}$                                 &$-$                                              &$-$                                            &$-0.0017\pm0.0021\pm0.0018$                        \\
        $\alpha^{\bar\Sigma^0}$                             &$-$                                              &$-$                                            &$0.0021\pm0.0020\pm0.0022$         \\
                            
        $\alpha^{\Sigma^+}_+$                               &$-$                                              &$0.0481\pm0.0031\pm0.0019$     &$-$                                                \\
        $\alpha^{\bar\Sigma^-}_-$                           &$-$                                              &$-0.0565\pm0.0047\pm0.0022$                    &$-$                                                \\
    
        $\alpha^{\Sigma^+}_+/\alpha^{\Sigma^+}_0$           &$-$                                              &$-0.0490\pm0.0032\pm0.0021$                    &$-$                                                \\
        $\alpha^{\bar\Sigma^-}_-/\alpha^{\bar\Sigma^-}_0$   &$-$                                              &$-0.0571\pm0.0053\pm0.0032$                    &$-$                                                \\
        
        $A_{CP}^{\Sigma}$                                   &$-0.0118\pm0.0083\pm0.0028$                      &$-0.080\pm0.052\pm0.028$                       &$0.0004\pm0.0029\pm0.0013$         \\
        $\alpha^{\Lambda}_{-}$                              &$-$                                              &$-$                                            &$0.730\pm0.051\pm0.011$            \\
        $\alpha^{\bar\Lambda}_{+}$                          &$-$                                              &$-$                                            &$-0.776\pm0.054\pm0.010$                           \\
        $A_{CP}^{\Lambda}$                                  &$-$                                              &$-$                                            &$-0.030\pm0.069\pm0.015$                           \\
      \bottomrule
        \end{tabular}}
    \label{tab:sigma}
\end{adjustwidth}
\end{table}

\textls[-15]{The additional weak decays of the $\Xi^-$ and $\Xi^0$ hyperons allow the extraction of their weak phases $\phi_{\Xi^-}$ and $\phi_{\Xi^0}$, as well as the strong and weak phase differences between the $S$-wave and $P$-wave amplitudes, thereby providing access to more $CP$ observables~\cite{BESIII:2021ypr,BESIII:2022lsz,BESIII:2023drj,BESIII:2023lkg,Liu:2023xhg,BESIII:2025dke}}. The corresponding polarization signals are shown in Figure~\ref{fig:polarization}d,e, and the results are summarized in Table~\ref{tab:xi}. Notably, the results for the $J/\psi$ and $\psi(3686)$ decays to $\Xi^0\bar{\Xi}^0$ exhibit a significant difference, which may provide an important probe into the decay dynamics of charmonium decays to hyperon pairs.
Similar to the case of $\Sigma^+$, a polarization measurement for single-sided neutron decays has been performed based on the process $J/\psi \to \Xi^-\bar{\Xi}^+$ $\to$ $\Lambda(p\pi^-)\pi^-$ $\bar{\Lambda}(\bar{n}\pi^0)\pi^+$ and its charge-conjugate channel~\cite{BESIII:2023jhj}. The decay parameters $\alpha^{\Lambda}_-$ (${\alpha}^{\bar\Lambda}_+$) and $\alpha^{\Lambda}_0$ (${\alpha}^{\bar\Lambda}_0$) of the two $\Lambda$ decay modes are measured separately. Also, the ratios $\alpha^{\Lambda}_0/\alpha^{\Lambda}_-$ and $\alpha^{\bar{\Lambda}}_{0}/\alpha^{\bar{\Lambda}}_{+}$, serving as observables for testing the $\Delta I = 1/2$ rule, are found to be smaller than the unity by more than $5\sigma$. The corresponding $CP$ observables $A^-_{CP}$ and $A^0_{CP}$ are used to test $CP$ symmetry. Furthermore, by taking the isospin-averaged value of the $\Lambda$ decays, $A^{\Lambda}_{CP} = (2A^-_{CP} + A^0_{CP})/3$, a more precise test of $CP$ symmetry can be achieved. The relevant results are also summarized in Table~\ref{tab:xi}.

\begin{figure}[H]

\begin{adjustwidth}{-\extralength}{0cm}
\centering 
{\includegraphics[width=1.0\textwidth]{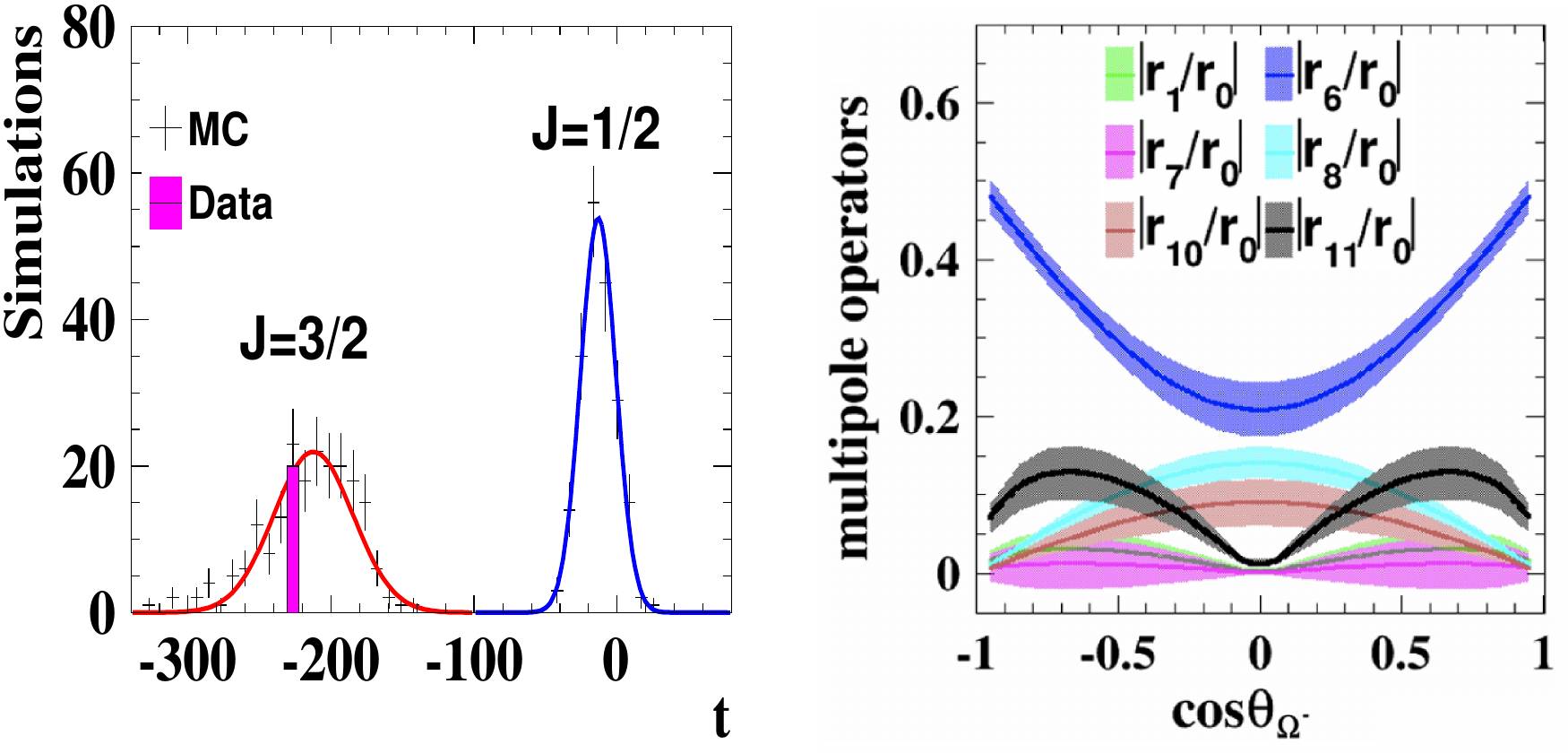}}
\end{adjustwidth}
\caption{{Distribution} 
 of the test statistic $t=S^{J=1/2}-S^{J=3/2}$ (\textbf{left}), where the curves denote Gaussian fits to the simulated samples and the vertical bar marks the value of $t$ obtained from data. The dependence of the multipolar polarization operators on $\cos\theta_{\Omega^-}$ (\textbf{right}), where the solid lines represent the central values, and the shaded areas represent the range within one standard deviation~\cite{BESIII:2020lkm}.}\label{fig:omega}
\end{figure}

The angular distribution analysis method can also be applied to high-spin hyperons. The discovery of the $\Omega^-$ was a major triumph for the eightfold way model of baryons, and angular distribution analysis provides a means to test the predictions of the eightfold way and the quark model regarding the spin of the $\Omega^-$.
Recently, the BESIII experiment performed a model-independent measurement of the $\Omega^-$ spin using the reaction $\psi(3686)\to\Omega^-\bar\Omega^+$ with the subsequent decays $\Omega^-\to K^-\Lambda$ and $\Lambda\to p\pi^-$~\cite{BESIII:2020lkm}. The $\Omega^-$ spin is determined to be $J= 3/2$, with a statistical significance of 14$\sigma$ over the $J= 1/2$ hypothesis, as shown in Figure~\ref{fig:omega} (left). This represents the first establishment of the $\Omega^-$ spin that is independent of any model-based assumptions since its discovery more than fifty years ago.
Additionally, the $\Omega^{-}$ particles exhibit not only vector polarization, but also quadrupole and octupole polarization contributions~\cite{Doncel:1972ez,Dubnickova:1992ii}. Accordingly, the helicity amplitudes and the $\cos\theta_{\Omega^{-}}$ dependence of the multipolar polarization operators were determined, as listed in Figure~\ref{fig:omega} (right) and Table~\ref{tab:omega}. 
Note that, since a single-tag method is adopted for event reconstruction in this analysis, multiple solutions exist~\cite{Zhang:2023box}.
In addition, based on the decays $\Omega^- \to \Xi^0\pi^-$ and $\Omega^- \to \Xi^-\pi^0$, the $\Delta I = 1/2$ rule can also be tested through the ratio of their branching fractions. A BESIII measurement reports $\mathcal{B}(\Omega^- \to \Xi^0\pi^-) = (25.03 \pm 0.44 \pm 0.53)\%$, $\mathcal{B}(\Omega^- \to \Xi^-\pi^0) = (8.43 \pm 0.52 \pm 0.28)\%$, and $\mathcal{B}(\Omega^- \to \Lambda K^-) = (66.3 \pm 0.8 \pm 2.0)\%$~\cite{BESIII:2023ldd}. The ratio of $\mathcal{B}(\Omega^- \to \Xi^0\pi^-)$ to $\mathcal{B}(\Omega^- \to \Xi^-\pi^0)$ is determined to be $2.97 \pm 0.19 \pm 0.11$, providing evidence for a deviation from the prediction of the $\Delta I = 1/2$ rule.

\begin{table}[H]
\caption{{Fitting} 
 results of each decay parameter for $J/\psi/\psi(3686)\to\Xi^{-}\bar\Xi^{+}$ and $J/\psi/\psi(3686)\to\Xi^{0}\bar\Xi^{0}$. 
		The first uncertainty is statistical and the second is systematic. The superscript or subscript $\Xi(\bar{\Xi})$ denotes $\Xi^-(\bar{\Xi}^+)$ or $\Xi^0(\bar{\Xi}^0)$, corresponding to the respective decay mode. $\alpha^{\Lambda}_{-}$ ($\alpha^{\bar\Lambda}_{+}$) denotes the decay parameter of the dominant decay mode $\Lambda \to p\pi^-$ ($\bar\Lambda \to \bar{p}\pi^+$), while $\alpha^{\Lambda}_{0}$ ($\alpha^{\bar\Lambda}_{0}$) represents that of the $\Lambda \to n\pi^0$ ($\bar\Lambda \to \bar{n}\pi^0$) decay.}
		\scriptsize
\begin{adjustwidth}{-\extralength}{0cm}
		\centering
\setlength{\tabcolsep}{1.9mm}{\begin{tabular}{lccccc}
\toprule
		\textbf{Parameter}			& \boldmath{$J/\psi\to\Xi^-\bar\Xi^+$}~\textbf{\cite{BESIII:2021ypr}} &\boldmath{$J/\psi\to\Xi^-\bar\Xi^+~(\Lambda \to n\pi^0)$}~\textbf{\cite{BESIII:2023jhj}} &\boldmath{$J/\psi\to\Xi^0\bar\Xi^0$}~\textbf{\cite{BESIII:2023drj}} & \boldmath{$\psi(3686)\to\Xi^-\bar\Xi^+$}~\textbf{\cite{BESIII:2022lsz}} & \boldmath{$\psi(3686)\to\Xi^0\bar\Xi^0$}~\textbf{\cite{BESIII:2025dke}}                 \\	
	\midrule
		$\alpha_{\psi}$ 				                  & $0.586 \pm 0.012 \pm 0.010$  &~~~$0.611\pm0.007^{+0.013}_{-0.007}$         &~~~$0.514  \pm 0.006 \pm  0.015$     &~~~$0.693 \pm 0.048 \pm 0.049$	  &~~~$0.768 \pm 0.029 \pm 0.025$       \\
		$\Delta{\Phi}$ (rad)			                  & $1.213 \pm 0.046 \pm 0.016$  &~~~$1.30\pm0.03^{+0.02}_{-0.03}$             &~~~$1.168  \pm 0.019 \pm  0.018$     &~~~$0.667 \pm 0.111 \pm 0.058$	  &~~~$0.257 \pm 0.061 \pm 0.009$       \\
		$\alpha_{\Xi}$			                          & $-0.376 \pm 0.007 \pm 0.003$                 &~~~$-0.367\pm0.004^{+0.003}_{-0.004}$                        &~~~$-0.3750 \pm 0.0034 \pm 0.0016$                   &~~~$-0.344 \pm 0.025 \pm 0.007$	                              &~~~$-0.345 \pm 0.015 \pm 0.003$                      \\
		$\alpha_{\bar{\Xi}}$		                      & $0.371 \pm 0.007 \pm 0.002$  &~~~$0.374\pm0.004^{+0.003}_{-0.004}$         &~~~$0.3790 \pm 0.0034 \pm 0.0021$    &~~~$0.355 \pm 0.025 \pm 0.002$	  &~~~$0.355 \pm 0.015 \pm 0.002$       \\
		$\phi_{\Xi}$ (rad)			                      & $0.011 \pm 0.019 \pm 0.009$  &~~~$-0.016\pm0.012^{+0.004}_{-0.008}$                        &~~~$0.0051 \pm 0.0096 \pm 0.0018$    &~~~$0.023 \pm 0.074 \pm 0.003$	  &~~~$0.008 \pm 0.041 \pm 0.002$       \\
 		$\phi_{\bar{\Xi}}$ (rad)	                      & $-0.021 \pm 0.019 \pm 0.007$	               &~~~$0.010\pm0.012^{+0.003}_{-0.013}$         &~~~$-0.0053 \pm 0.0097 \pm 0.0019$                   &~~~$-0.123 \pm 0.073 \pm 0.004$	                          &~~~$-0.009 \pm 0.040 \pm 0.004$                      \\ 
 		$\alpha^{\Lambda}_{-}$		                        & $0.757  \pm 0.011 \pm 0.008$ &~~~$0.764\pm0.008^{+0.005}_{-0.006}$         &~~~$0.7551 \pm 0.0052 \pm 0.0023$    &~~~$-$                                               &~~~$-$                                               \\ 
        $\alpha^{\bar\Lambda}_{+}$                          & $-0.763 \pm 0.011 \pm 0.007$                 &~~~$-0.774\pm0.009^{+0.005}_{-0.005}$                        &~~~$-0.7448 \pm 0.0052 \pm 0.0017$                   &~~~$-$                                               &~~~$-$                                               \\          
        $\alpha^{\Lambda}_{0}$		                        & $-$                                          &~~~$0.670\pm0.009^{+0.009}_{-0.008}$         &~~~$-$                                               &~~~$-$                                               &~~~$-$                                               \\ 
        $\bar\alpha^{\Lambda}_{0}$                          & $-$                                          &~~~$-0.668\pm0.008^{+0.006}_{-0.008}$                        &~~~$-$                                               &~~~$-$                                               &~~~$-$                                               \\
        $\alpha^{\Lambda}_{0}/\alpha^{\Lambda}_{-}$         & $-$                                          &~~~$0.877\pm0.015^{+0.014}_{-0.010}$         &~~~$-$                                               &~~~$-$		                                         &~~~$-$                                               \\
        $\alpha^{\bar\Lambda}_{0}/\alpha^{\bar\Lambda}_{+}$ & $-$                                          &~~~$0.863\pm0.014^{+0.012}_{-0.008}$         &~~~$-$                                               &~~~$-$		                                         &~~~$-$                                               \\
		$A_{CP}^{\Xi}$ ($\times10^{-2}$) 					& $0.60 \pm 1.34 \pm 0.56$     &~~~$-0.9\pm0.8^{+0.7}_{-0.2}$                                &~~~$-0.54 \pm 0.65 \pm 0.31$                         &~~~$-1.5 \pm 5.1 \pm 1.0$		                     &~~~$-0.014\pm0.030\pm0.010$                          \\
		$A_{CP}^{-}$ ($\times10^{-3}$)                      & $-4   \pm 12   \pm 9$                        &~~~$-7\pm8^{+2}_{-3}$                                        &~~~$6.9 \pm 5.8 \pm 1.8$             &~~~$-$		                                         &~~~$-$                                               \\
        $A_{CP}^{0}$ ($\times10^{-3}$)                      & $-$                                          &~~~$1\pm9^{+5}_{-7}$                         &~~~$-$                                               &~~~$-$		                                         &~~~$-$                                               \\
        $A_{CP}^{\Lambda}$ ($\times10^{-3}$)                & $-$                                          &~~~$-4\pm7^{+3}_{-4}$                                        &~~~$-$                                               &~~~$-$		                                         &~~~$-$                                               \\
		$\Delta\phi_{CP}^{\Xi}$ ($\times10^{-2}$ rad)		& $-0.48 \pm 1.37 \pm 0.29$                    &~~~$-0.3\pm0.8^{+0.3}_{-0.7}$                                &~~~$-0.1 \pm 6.9 \pm 0.9$                            &~~~$-5.0 \pm 5.2 \pm 0.3$		                     &~~~$0.000\pm0.028\pm0.003$           \\
		$\delta_{p}-\delta_{s}$ ($\times10^{-1}$ rad)	    & $-0.40 \pm 0.33 \pm 0.17$                    &~~~$0.33\pm0.20^{+0.08}_{-0.12}$             &~~~$-0.13 \pm 0.17 \pm 0.4$                          &~~~$-2.0 \pm 1.3 \pm 0.1$		                     &~~~$-0.022\pm0.078\pm0.011$                          \\
		$\xi_{p}-\xi_{s}$ ($\times10^{-1}$ rad)	            & $0.12 \pm 0.34 \pm 0.08$     &~~~$0.07\pm0.20^{+0.18}_{-0.05}$             &~~~$0.00\pm 0.17 \pm 0.02$           &~~~$-$		                               &~~~$0.001\pm0.076\pm0.011$           \\
	\bottomrule
		\end{tabular}}
	\label{tab:xi}
\end{adjustwidth}
\end{table}
\vspace{-15pt}

\begin{table}[H]
    \caption{{Summary} 
 of the fitted values of helicity parameters in $\psi(3686)\to\Omega^-\bar\Omega^+$ with the spin-3/2 hypothesis~\cite{BESIII:2020lkm}. The first uncertainties are statistical, and the second ones are systematic.}
    \label{tab:omega}
     \centering
\setlength{\tabcolsep}{11.4mm}{\begin{tabular}{ccc}
       \toprule
        \textbf{Parameter}       & \textbf{Solution I}               & \textbf{Solution II}  \\ 
        \midrule
        $h_1$           & $0.30 \pm 0.11 \pm 0.04$ & $0.31 \pm 0.10 \pm 0.04$ \\
        $\phi_1$        & $0.69 \pm 0.41 \pm 0.13$ & $2.38 \pm 0.37 \pm 0.13$ \\
        $h_3$           & $0.26 \pm 0.05 \pm 0.02$ & $0.27 \pm 0.05 \pm 0.01$ \\
        $\phi_3$        & $2.60 \pm 0.16 \pm 0.08$ & $2.57 \pm 0.16 \pm 0.04$ \\
        $h_4$           & $0.51 \pm 0.03 \pm 0.01$ & $0.51 \pm 0.03 \pm 0.01$ \\
        $\phi_4$        & $0.34 \pm 0.80 \pm 0.31$ & $1.37 \pm 0.68 \pm 0.16$ \\
        $\phi_{\Omega}$ & $4.29 \pm 0.45 \pm 0.23$ & $4.15 \pm 0.44 \pm 0.16$ \\
      \bottomrule
    \end{tabular}}
\end{table}
\unskip

\subsection{$e^+e^-$ Annihilation}
According to Equation~(\ref{eq:eff}), the angular distribution parameter is related to the moduli of the EMFFs. Using the measured angular distribution parameter (denoted as $\eta$ for the non-resonant process), the modulus ratio $R$ of the EMFFs can be directly calculated as~\cite{Wang:2022zyc}
\begin{equation}
    R = \frac{|G_E|}{|G_M|} = \sqrt{\frac{\tau(1-\eta)}{1+\eta}},
\end{equation}
where $\tau = s/4M^2$, and $M$ is the hyperon mass. Variations in the value of $R$ correspond to different production mechanisms of hyperons.
To investigate the production mechanisms and internal structure of hyperons under different interactions, BESIII performed angular distribution analyses of $\Lambda$ and $\Sigma^+$ hyperon pairs at energies near the threshold, around the $\psi(3686)$ and $\psi(3770)$ resonances, covering both purely electromagnetic interactions and mixed electromagnetic–strong interactions.

\begin{figure}[H]
 
\includegraphics[width=1.0\textwidth]{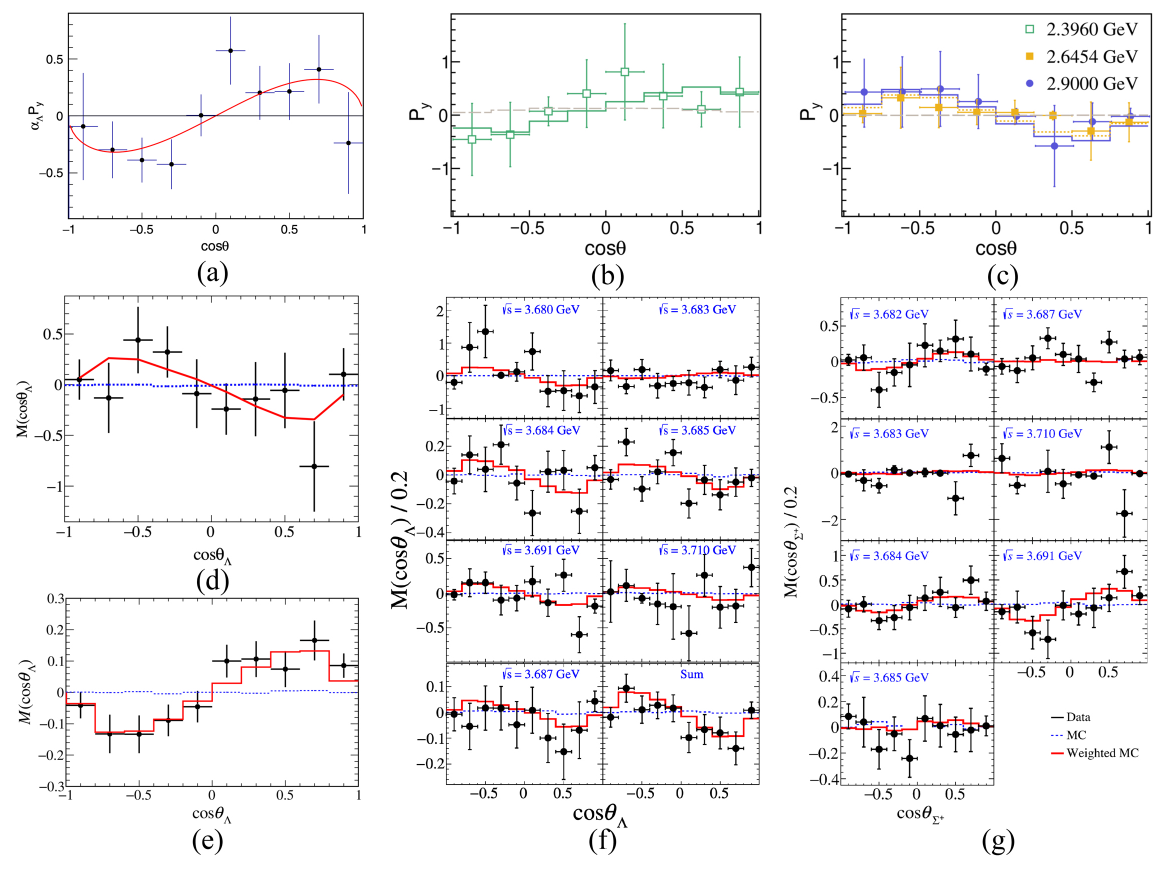}
\caption{{Distributions} 
 of the polarization observables $P_y$ for $e^+e^- \to \Lambda\bar{\Lambda}$ at $\sqrt{s} = 2.396$ GeV~\cite{BESIII:2019nep}~(\textbf{a}) \textls[-25]{and $e^+e^- \to \Sigma^+\bar{\Sigma}^-$ at $\sqrt{s} = 2.396$–$2.900$ GeV~\cite{BESIII:2023ynq} (\textbf{b},\textbf{c}), and $M$ for $e^+e^- \to \Lambda\bar{\Lambda}$ at $\sqrt{s} = 3.773$~GeV~\cite{BESIII:2023euh,BESIII:2025yzk}} {(\textbf{d},\textbf{e})} and $3.68$–$3.71$ GeV~\cite{BESIII:2021cvv} (\textbf{f}) and $e^+e^- \to \Sigma^+\bar{\Sigma}^-$ at $\sqrt{s} = 3.68$–$3.71$ GeV~\cite{BESIII:2024dmr} (\textbf{g}), as functions of $\cos\theta$ for each process. The dots with error bars represent the experimental data, the red curve shows the global fit, the gray and blue dashed histograms correspond to the PHSP MC, and the red polyline and colored solid histograms show the PHSP MC weighted according to the global fit.}
\label{fig:FF}
\end{figure}

For the $\Lambda$ hyperon, a study of the purely electromagnetic process was performed at $\sqrt{s}=2.396$ GeV~\cite{BESIII:2019nep}, and the distribution of the polarization observable is shown in Figure~\ref{fig:FF}a. Measurements near the $\psi(3686)$~\cite{BESIII:2021cvv} and $\psi(3770)$~\cite{BESIII:2023euh,BESIII:2025yzk} resonances are shown in Figure~\ref{fig:FF}d--f. 
For the $\Sigma^{+}$ hyperon, studies are currently available only near the threshold~\cite{BESIII:2023ynq} and around the $\psi(3686)$~\cite{BESIII:2024dmr} resonance. The corresponding polarization signals are shown in Figure~\ref{fig:FF}b,c,g. All of the above measurement results are summarized in Table~\ref{tab:FF}. These studies provide valuable insights into the production mechanisms of $\Lambda$ and $\Sigma^{+}$ hyperon pairs at different c.m. energies.

\begin{table}[H]
    \caption{The measured parameters of $e^+e^-\to\Lambda\bar\Lambda$~\cite{BESIII:2019nep,BESIII:2021cvv,BESIII:2025yzk} and $e^+e^-\to\Sigma^+\bar\Sigma^-$~\cite{BESIII:2023ynq,BESIII:2024dmr} at each energy point. The first uncertainty is statistical and the second is systematic.}
   \footnotesize
  \setlength{\tabcolsep}{5.05mm}{\begin{tabular}{l c c c}
       \toprule
        \boldmath{$\sqrt{s}$} \textbf{(GeV)} 	       & \boldmath{$\eta$}     & \boldmath{$\Delta\Phi$} \textbf{(}\boldmath{$^\circ$}\textbf{)}      & \boldmath{$R$}       \\
      \midrule
        \multicolumn{4}{c}{$e^+e^-\to\Lambda\bar\Lambda$} \\
        2.396                      & $0.12\pm0.14\pm0.02$           & $37\pm12\pm6$                                    & $0.96\pm0.14\pm0.02$          \\
        3.68--3.71                & $0.69^{+0.07}_{-0.07}\pm0.02$  & $23^{+8.8}_{-8.0}\pm1.6$                         & $0.71^{+0.10}_{-0.10}\pm0.03$ \\    
        3.773                      & $0.86\pm0.05\pm0.03$           & $88\pm21\pm2$                                    & $0.47\pm0.08\pm0.05$          \\[3pt]
        \multicolumn{4}{c}{$e^+e^-\to\Sigma^+\bar\Sigma^-$} \\
        2.396                      & $-0.47\pm0.18\pm0.09$          & $-42\pm22\pm14~(-138\pm22\pm14)$                 & $1.69\pm0.38\pm0.20$          \\
        2.645                      & $0.41\pm0.12\pm0.06$           & $55\pm19\pm14$                                   & $0.72\pm0.11\pm0.06$          \\
        2.900                      & $0.35\pm0.17\pm0.15$           & $78\pm22\pm9$                                    & $0.85\pm0.16\pm0.15$          \\
        3.682                      & $0.54\pm0.17\pm0.12$           & $22 \pm 23 \pm 7$                                & $0.84\pm0.20\pm0.14$          \\
        3.683                      & $0.96\pm0.13\pm0.12$           & $135 \pm 95 \pm 7$                               & $0.22\pm0.37\pm0.34$          \\
        3.684                      & $0.86\pm0.15\pm0.12$           & $68 \pm 35 \pm 7$                                & $0.42\pm0.21\pm0.20$          \\
        3.685                      & $0.76\pm0.10\pm0.12$           & $9 \pm 18 \pm 7$                                 & $0.57\pm0.15\pm0.16$          \\
        3.687                      & $0.66\pm0.12\pm0.12$           & $1 \pm 15 \pm 7$                                 & $0.70\pm0.15\pm0.15$          \\
        3.691                      & $0.16\pm0.23\pm0.12$           & $74 \pm 31 \pm 7$                                & $1.31\pm0.30\pm0.16$          \\
        3.710                      & $0.01\pm0.40\pm0.12$           & $-151 \pm 34 \pm 7$                              & $1.55\pm0.62\pm0.19$          \\
    \bottomrule
        \end{tabular}}
    \label{tab:FF}
\end{table}

\subsection{Electric Dipole Moment}
Furthermore, if we introduce four form factors, $F_V$, $H_{\sigma}$, $F_A$, and $H_T$, representing the Dirac, Pauli, parity violation, and electric dipole form factors, respectively, the amplitude of the hyperon pair production process can be written as follows~\cite{He:2022jjc,Fu:2023ose}:
\begin{equation}
    \mathcal{A}^{\mu} = \bar{u}(p_1)[\gamma^{\mu}F_V + \frac{i}{2M}\sigma^{\mu\nu}q_{\nu}H_{\sigma} + \gamma^{\mu}\gamma^5F_A + \sigma^{\mu\nu}q_{\nu}\gamma^5H_T]v(p_2),
\end{equation}
where $p_1$, $p_2$, and $M$ follow the same definitions as in Equation~(\ref{vtx}), Parity conservation implies $F_A = 0$, while $CP$ conservation further implies $H_T = 0$. The form factors $F_V$ and $H_{\sigma}$ are related to the EMFFs with $F_V = G_M - 4M^2\frac{G_M - G_E}{(p_1 - p_2)^2}$ and $H_{\sigma} = 4M^2\frac{G_M - G_E}{(p_1 - p_2)^2}$, while $H_T$ can be associated with the hyperon electric dipole moment (EDM)~\cite{He:2022jjc,Fu:2023ose}. Based on this approach, a joint angular distribution can be constructed to explore $CP$ violation in a novel way, while simultaneously determining the hyperon EDM.
Recently, BESIII performed an analysis based on the $J/\psi \to \Lambda \bar\Lambda$ process, using $F_A$ and $H_T$ as observables to explore both $P$ and $CP$ violation, while also making a precise measurement of the EDM of $\Lambda$~\cite{BESIII:2025vxm}. The corresponding results are shown in Table~\ref{tab:edm}.
This measurement improves the sensitivity to the $\Lambda$ EDM by three orders of magnitude compared to previous results, while also introducing a new approach to probing possible sources of $CP$ violation.
\begin{table}[H]	
    \caption{Summary of fitted parameters~\cite{BESIII:2025vxm}. The first uncertainty is statistical and the second is~systematic.}
  
\setlength{\tabcolsep}{16.4mm}{\begin{tabular}{l l}
     \toprule
        \textbf{Parameter} 	                        &  \textbf{Fitting Results}           \\
     \midrule
        $P_L$                               & ~~~~$(-1.8 \pm 1.2 \pm 0.8) \times 10^{-3}$       \\
        $Re(F_A)$                           & ~~~~$(-2.4 \pm 1.6 \pm 3.1) \times 10^{-6}$       \\
        $Im(F_A)$                           & ~~~~$(-7.9 \pm 3.7 \pm 2.5) \times 10^{-6}$       \\
        $Re(H_T)~({\rm GeV}^{-1})$          & ~~~~$(-1.4 \pm 1.4 \pm 0.2) \times 10^{-6}$       \\
        $Im(H_T)~({\rm GeV}^{-1})$          & ~~~~$(1.3 \pm 1.2 \pm 0.4) \times 10^{-6}$        \\
        $Re(d_{\Lambda})~({\rm cm})$        & ~~~~$(-3.1 \pm 3.2 \pm 0.5) \times 10^{-19}~e$    \\
        $Im(d_{\Lambda})~({\rm cm})$        & ~~~~$(2.9 \pm 2.6 \pm 0.6) \times 10^{-19}~e$     \\
       \bottomrule
        \end{tabular}}
    \label{tab:edm}
\end{table}


\section{Hyperon Rare Decay}
\subsection{Radiative Decay}
The radiative decays of hyperons offer important insights into the properties of nonleptonic weak interactions~\cite{Behrends:1958zz}. In the limit of unitary symmetry, the parity-violating amplitude in these decays is expected to be small, resulting in a decay asymmetry parameter of $\alpha_{\gamma} = 0$~\cite{Hara:1964zz}. However, experimental measurements show that the values of $\alpha_\gamma$ are large. Several phenomenological models have been proposed to account for these results~\cite{Gavela:1980bp,Nardulli:1987ub,Zenczykowski:1991mx,Borasoy:1999nt,Zenczykowski:2005cs,Niu:2020aoz,Shi:2022dhw}; nevertheless, none has yet provided a unified description of all weak radiative hyperon decays. To enrich the available experimental information, BESIII has also carried out a series of studies on radiative hyperon decays.

\begin{figure}[H]
 
\includegraphics[width=0.9\textwidth]{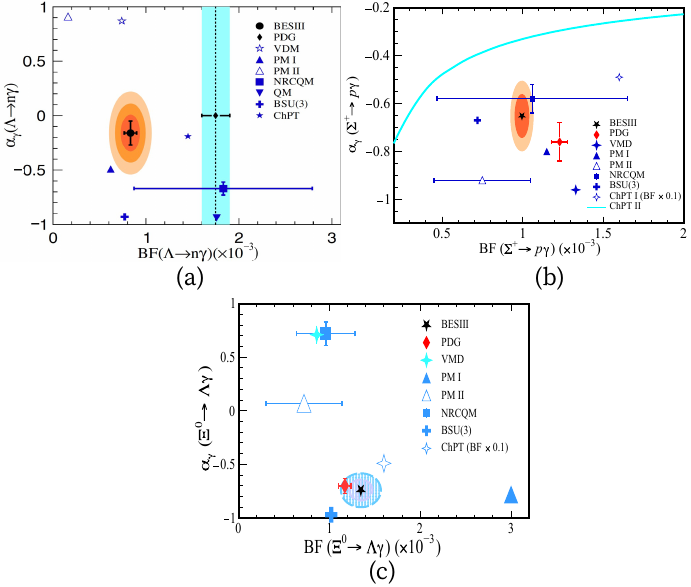}
\caption{\label{fig:rad_dec} Distributions of $\alpha_{\gamma}$ versus branching fraction for $\Lambda \to n\gamma$~\cite{BESIII:2022rgl} (\textbf{a}), $\Sigma^+p\gamma$~\cite{BESIII:2023fhs} (\textbf{b}), and $\Xi^0\to\Lambda\gamma$~\cite{BESIII:2024lio} (\textbf{c}), along with the PDG values~\cite{ParticleDataGroup:2024cfk} and predictions from the vector meson dominance model (VDM)~\cite{Zenczykowski:1991mx}, pole model (PM I~\cite{Gavela:1980bp} and PM II~\cite{Nardulli:1987ub}), nonrelativistic constituent quark model (NRCQM)~\cite{Niu:2020aoz}, broken SU(3) model [BSU(3)]~\cite{Zenczykowski:2005cs}, and chiral perturbation theory (ChPT)~\cite{Borasoy:1999nt,Shi:2022dhw}.}
\end{figure}

Based on the processes $J/\psi \to \Lambda\bar{\Lambda}$, $\Sigma^+\bar{\Sigma}^-$, and $\Xi^0\bar{\Xi}^0$, the branching fractions and decay asymmetry parameters of the decays $\Lambda \to n\gamma$, $\Sigma^+ \to p\gamma$, and $\Xi^0 \to \Lambda\gamma$ were measured~\cite{BESIII:2022rgl,BESIII:2023fhs,BESIII:2024lio}. Compared with previous results, the precision of these measurements has been significantly improved, and the decay asymmetry parameter of $\Lambda \to n\gamma$ has been determined for the first time. The measured results are summarized in Table~\ref{tab:rad_dec}. These measurements serve as valuable tests of various phenomenological models. A comparison between the BESIII results, the PDG averages, and the model predictions is presented in Figure~\ref{fig:rad_dec}. In addition, $CP$ tests were performed in these analyses by constructing observables from the $\alpha_\gamma$ values of hyperons and antihyperons. Within the current statistical precision, the results are consistent with $CP$ symmetry.
 
\begin{table}[H]
    \caption{{Results} 
 of branching fractions and $\alpha_{\gamma}$ for each radiative decay obtained from both individual and simultaneous fits~\cite{BESIII:2022rgl,BESIII:2023fhs,BESIII:2024lio}. The first uncertainties are statistical and the second uncertainties are~systematic.} 
    
\setlength{\tabcolsep}{7.3mm}
 \resizebox{\linewidth}{!}{\begin{tabular}{l c c }
       \toprule
        Mode                            & $\Lambda \to n\gamma$         &$\bar\Lambda \to \bar{n}\gamma$    \\
       \midrule
        Individual BF ($10^{-3}$)       & $0.820\pm0.045\pm0.066$       & $0.862\pm0.071\pm0.084$           \\
        Simultaneous BF ($10^{-3}$)     & \multicolumn{2}{c}{$0.832\pm0.038\pm0.054$}                       \\ 
        Individual $\alpha_{\gamma}$    & $-0.13\pm0.13\pm0.03$         & $0.21\pm0.15\pm0.06$              \\
        Simultaneous $\alpha_{\gamma}$  & \multicolumn{2}{c}{$-0.16\pm0.010\pm0.05$}                        \\
    \midrule
        Mode                            & $\Sigma^+ \to p\gamma$        &$\bar\Sigma^- \to \bar{p}\gamma$   \\
        \midrule
        Individual BF ($10^{-3}$)       & $1.005\pm0.032$               & $0.993\pm0.030$                   \\
        Simultaneous BF ($10^{-3}$)     & \multicolumn{2}{c}{$0.996\pm0.021\pm0.018$}                       \\ 
        Individual $\alpha_{\gamma}$    & $-0.587\pm0.082$              & $0.710\pm0.076$                   \\
        Simultaneous $\alpha_{\gamma}$  & \multicolumn{2}{c}{$-0.651\pm0.056\pm0.020$}                      \\   
     \midrule
        Mode                            & $\Xi^0 \to \Lambda\gamma$     &$\bar\Xi^0 \to \bar\Lambda\gamma$  \\
       \midrule
        Individual BF ($10^{-3}$)       & $1.348\pm0.090\pm0.054$       & $1.326\pm0.098\pm0.066$           \\
        Simultaneous BF ($10^{-3}$)     & \multicolumn{2}{c}{$1.347\pm0.066\pm0.054$}                       \\ 
        Individual $\alpha_{\gamma}$    & $-0.652\pm0.092\pm0.016$      & $0.830\pm0.080\pm0.044$           \\
        Simultaneous $\alpha_{\gamma}$  & \multicolumn{2}{c}{$-0.741\pm0.062\pm0.019$}                      \\ 
     \bottomrule
        \end{tabular}}
    \label{tab:rad_dec}
\end{table}

\subsection{Semileptonic Decay}
The general approach to determining the absolute branching fractions of rare hyperon decays is to extract the signal yields and detection efficiencies using both the single-tag (ST) and double-tag (DT) methods, and then to obtain the branching fraction as
\begin{equation}
    \mathcal{B}_{sig} = \frac{N_{DT/\epsilon_{DT}}}{N_{ST}/\epsilon_{ST}},
\end{equation}
where $N_{ST}$ ($N_{DT}$) denotes the ST (DT) yield, and $\epsilon_{ST}$ ($\epsilon_{DT}$) denotes the corresponding ST (DT) detection efficiency.
Based on this method, a study using the process $J/\psi \to \Xi^0 \bar{\Xi}^0$ was carried out at BESIII to search for the baryon and lepton number violating decays $\Xi^0 \to K^- e^+$ and $\Xi^0 \to K^+ e^-$~\cite{BESIII:2023str}. 
Since no evidence for either decay was observed in this analysis, upper limits on their branching fractions at the 90\% confidence level (C.L.) were reported to be $\mathcal{B}(\Xi^0 \to K^-e^+) < 3.6\times10^{-6}$ and $\mathcal{B}(\Xi^0 \to K^+e^-) < 1.9\times10^{-6}$.

BESIII has also investigated the semileptonic decays of the $\Lambda$~\cite{BESIII:2021ynj}, $\Xi^-$~\cite{BESIII:2021emv}, and $\Xi^0$~\cite{BESIII:2022alf} hyperons. 
The decays $\Lambda \to p \mu^- \bar{\nu}_\mu$ and $\Lambda \to p e^- \bar{\nu}_e$ provide an excellent opportunity to test lepton flavor universality (LFU) and thereby to examine the validity of the SM~\cite{Bifani:2018zmi}. The ratio of the decay rates of these two processes $R^{\mu e} \equiv [\Gamma(\Lambda \to p \mu^- \bar{\nu}_\mu)]/[\Gamma(\Lambda \to p e^- \bar{\nu}_e)]$ serves as an observable for testing LFU. However, previous results are still based on fixed-target experiments and therefore suffer from relatively large uncertainties.
Using a data sample of $10$ billion $J/\psi$ events and the decay $J/\psi \to \Lambda \bar{\Lambda}$, BESIII has measured the absolute branching fraction of $\Lambda \to p \mu^- \bar{\nu}_\mu$ to be $\mathcal{B}(\Lambda \to p \mu^- \bar{\nu}_\mu)=(1.48\pm0.21\pm0.08)\times10^{-4}$, achieving a 30\% improvement in precision compared to the world average.
Based on the well-measured branching fraction of $\Lambda \to p e^-\bar{\nu}_e$, the ratio $R^{\mu e}$ is determined to be $0.178 \pm 0.028$, which is consistent with previous results and with LFU~\cite{ParticleDataGroup:2024cfk,Chang:2014iba}.
Meanwhile, a $CP$ test is also performed using the asymmetry $A_{CP} = \frac{\mathcal{B}(\Lambda \to p \mu^- \bar{\nu}_\mu) - \mathcal{B}(\bar\Lambda \to \bar{p} \mu^+ \nu_\mu)}{\mathcal{B}(\Lambda \to p \mu^- \bar{\nu}_\mu) + \mathcal{B}(\bar\Lambda \to \bar{p} \mu^+ \nu_\mu)}$, which is determined to be $A_{CP} = 0.02\pm0.14\pm0.02$, consistent with $CP$ conservation.
In addition, using the same $J/\psi$ decay sample, the processes $\Xi^- \to \Xi^{0} e^- \bar{\nu}_e$ and $\Xi^{0} \to \Sigma^- e^+ \nu_e$ have also been investigated at BESIII. 
Since no significant signal is observed, upper limits on their branching fractions at 90\% C.L. are reported, given by $\mathcal{B}(\Xi^-\to\Xi^0 e^- \bar{\nu}_e) < 2.59\times10^{-4}$ and $\mathcal{B}(\Xi^0\to\Sigma^- e^+ \nu_e) < 1.6\times10^{-4}$.
These results also provide important experimental constraints for the study of the SU(3) symmetry-breaking mechanism, $\Delta S = \Delta Q$-violating decays, and other rare or forbidden hyperon decays.

\section{Hyperon Pair Production}
\begin{figure}[H]
 
\includegraphics[width=1.0\textwidth]{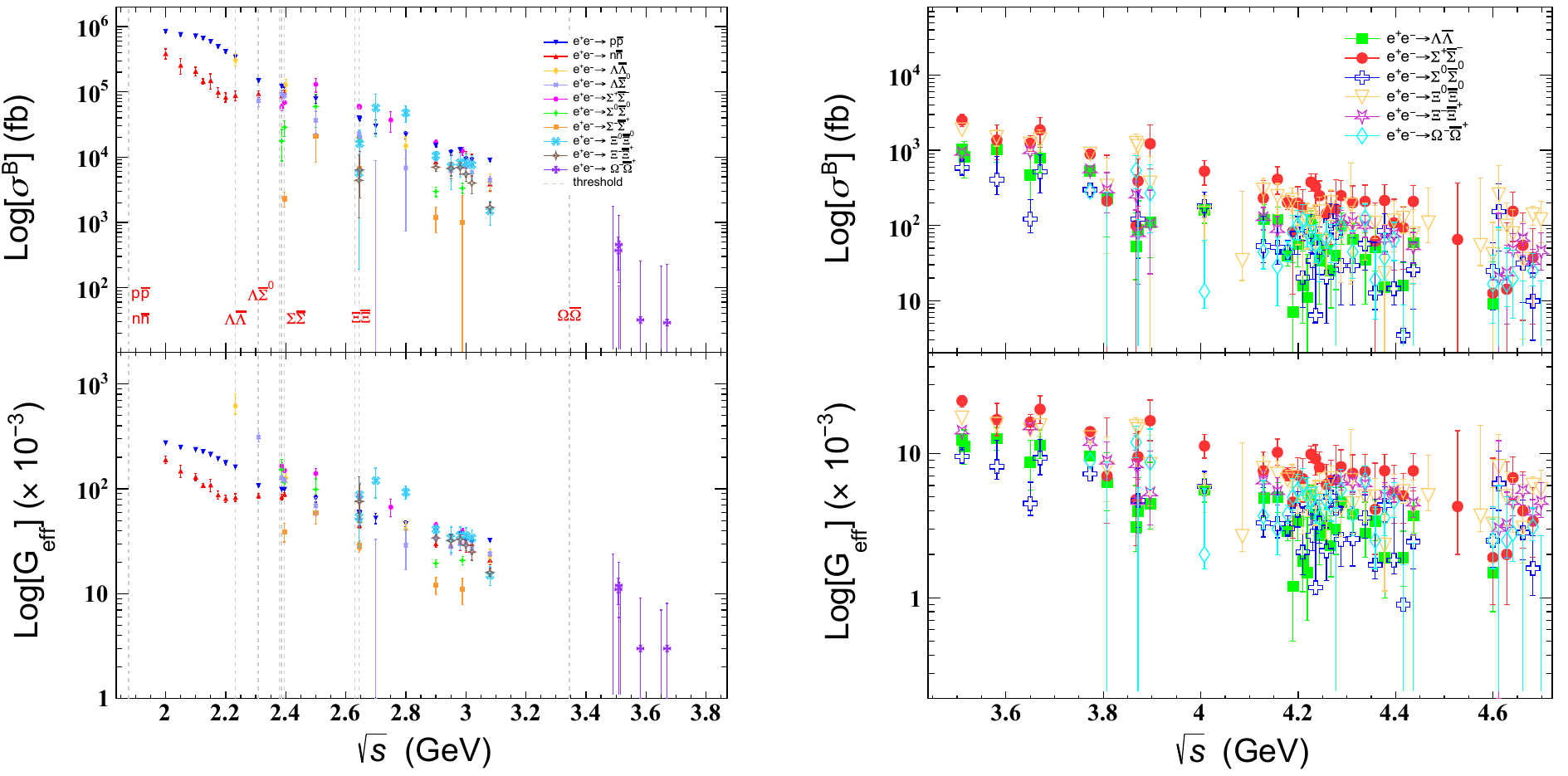}
\caption{The Born cross sections and effective form factors for each baryon pair process near the threshold (\textbf{left}) and in the charmonium energy region (\textbf{right}). The gray dashed lines represent the production threshold of corresponding baryon pairs.}\label{fig:cs_line}
\end{figure}

In recent years, measurements of cross sections near production thresholds have provided a sensitive probe for exploring particle interactions, electromagnetic form factors, and possible new resonance states. Numerous experimental studies have been carried out to explore baryon properties, with most analyses based on the interpretation of the Born cross section for baryon–antibaryon pair production. Notably, several of these measurements observed an unusual behavior near the threshold---a significantly nonzero cross section, commonly referred to as threshold enhancement. 
At BESIII, threshold enhancement has been observed in the measurements of the processes $e^+e^- \to p\bar{p}$~\cite{BESIII:2015axk,BESIII:2019tgo,BESIII:2019hdp}, $e^+e^- \to n\bar{n}$~\cite{BESIII:2021tbq}, $e^+e^- \to \Lambda\bar{\Lambda}$~\cite{BESIII:2017hyw}, and $e^+e^- \to \Lambda\bar{\Sigma}^0$~\cite{BESIII:2023pfv}. Similar threshold behavior has also been observed in the cross sections of $e^+e^-\to\Sigma\bar{\Sigma}$~\cite{BESIII:2020uqk,BESIII:2021rkn} and $e^+e^-\to\Xi\bar{\Xi}$~\cite{BESIII:2020ktn,BESIII:2021aer}, although larger data samples are required to draw more definitive conclusions. In addition, measurements of $e^+e^-\to\Omega^-\bar{\Omega}^+$ have also been performed~\cite{BESIII:2022kzc}. However, since the available energy points are relatively far from the production threshold, no evident threshold enhancement has been observed in this case.
In these studies, the Born cross sections and effective form factors of each process were measured using either energy-scan or ISR methods, as shown in Figure~\ref{fig:cs_line} (left). In the measurements of the reactions of $e^+e^-\to p\bar{p}$ and $n\bar{n}$, oscillatory behavior of the effective form factors has also been observed, suggesting the presence of intrinsic dynamics that are not yet fully understood.

\begin{figure}[H]

\includegraphics[width=1.0\textwidth]{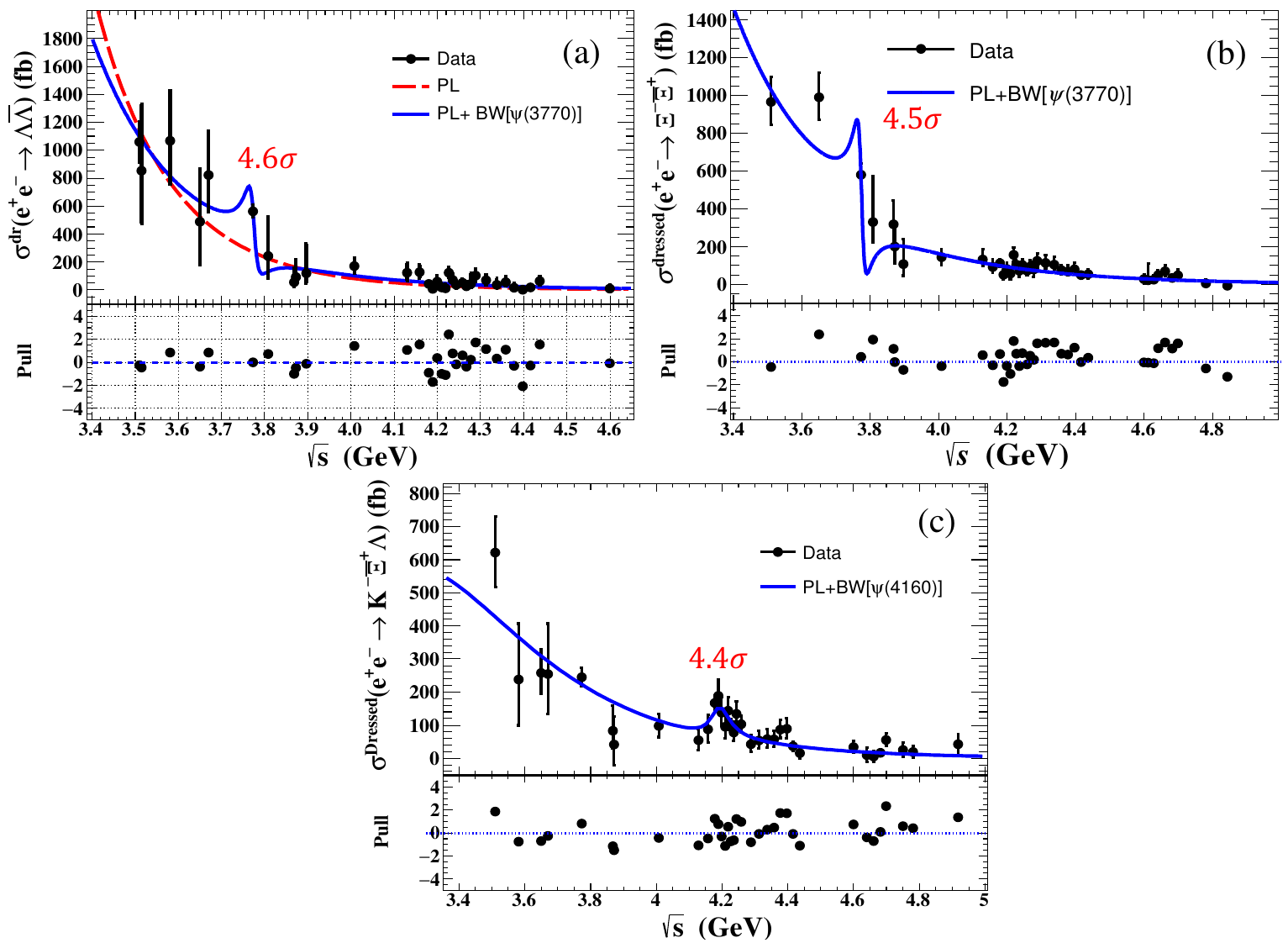}
\caption{{Fits} 
 to the dressed cross sections for the $e^+e^- \to \Lambda\bar{\Lambda}$~\cite{LamLam} (\textbf{a}), $\Xi^-\bar{\Xi}^+$~\cite{XiXiimp} (\textbf{b}), and $K^-\bar\Xi^+\Lambda$~\cite{BESIII:2024ogz} (\textbf{c}) reactions.}
\label{fig:cs_charm}
\end{figure}

In the charmonium energy region, studying the hadronic decays of charmonium-like states by measuring the Born cross sections of baryonic final states is one of the key approaches to revealing their physical properties. 
The BESIII experiment performed its first search for the charmonium(-like) states $Y(4230/4260)$ in the process $e^+e^- \to \Xi^-\bar{\Xi}^+$ at c.m. energies ranging from 3.5 to 4.6 GeV~\cite{BESIII:2019cuv}, but found no significant evidence for $Y(4230/4260)$ decaying into $\Xi^-\bar{\Xi}^+$. 
Subsequently, BESIII has conducted a series of studies on hyperon pair production in the processes $e^+e^- \to \Lambda\bar{\Lambda}$, $\Sigma^+\bar{\Sigma}^-$, $\Sigma^0\bar{\Sigma}^0$, $\Xi^0\bar{\Xi}^0$, and $\Omega^-\bar\Omega^+$,  along with an updated measurement of $\Xi^-\bar{\Xi}^+$~\cite{LamLam,SigSigc,Sig0Sig0,Xixi0,XiXiimp,BESIII:2025fph,Zhang:2025qmo}. The Born cross sections and effective form factors for these processes were measured, as shown in Figure~\ref{fig:cs_line} (right).
To further investigate the charmonium(-like) states, fits to the dressed cross sections were carried out. Evidence for the decays $\psi(3770)\to\Lambda\bar{\Lambda}$ and $\psi(3770)\to\Xi^-\bar{\Xi}^+$ was found~\cite{LamLam,XiXiimp}, as shown in Figure~\ref{fig:cs_charm}a,b. The branching fractions of these two decays are determined to be $\mathcal{B}(\psi(3770)\to\Lambda\bar{\Lambda})=(2.4^{+15.0}_{-1.9})\times10^{-5}$ or $(14.4^{+2.7}_{-14.0})\times10^{-5}$ and $\mathcal{B}(\psi(3770)\to\Xi^-\bar{\Xi}^+)=(136.0\pm35.2)\times10^{-6}$, respectively. For the other processes, no significant signal of charmonium(-like) states was observed. Only upper limits for the product of the branching fraction and the electronic partial width, $\Gamma_{ee}\mathcal{B}$, at the 90\% C.L., were evaluated. These measurements provide evidence for charmless decays of $\psi(3770)$ and can be useful for understanding the coupling of charmonium(-like) states to hyperon pair final states.

\begin{figure}[H]
 
\includegraphics[width=0.49\textwidth]{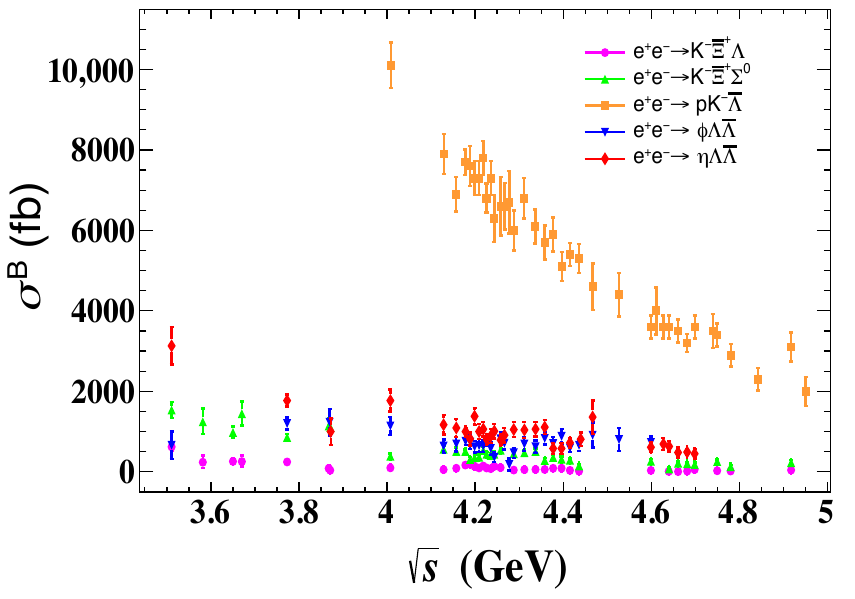}
\includegraphics[width=0.49\textwidth]{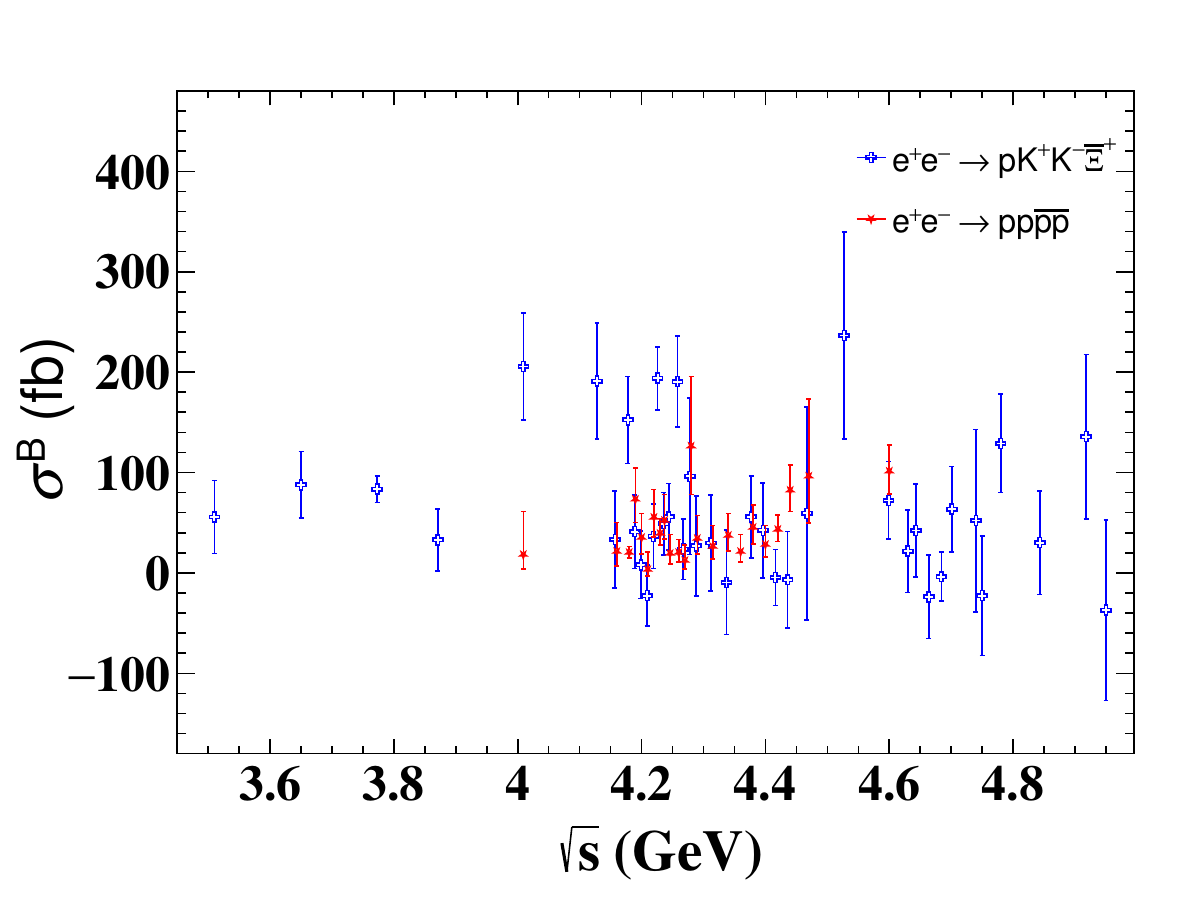}
\caption{{Line} 
 shapes of the measured Born cross sections for each three-body (\textbf{left}) and four-body (\textbf{right}) process.}\label{fig:cs_34body}
\end{figure}

The BESIII experiment has also measured the Born cross sections for three-body decays with hyperon final states~\cite{BESIII:2021fqx,BESIII:2022tvj,BESIII:2023ojh,BESIII:2023kgz,BESIII:2024ogz}, as shown in Figure~\ref{fig:cs_34body} (left). For these processes, no significant charmonium(-like) states were observed, except for evidence of the decay $\psi(4160) \to K^-\bar{\Xi}\Lambda + \text{c.c.}$, which was observed with a significance of 4.4$\sigma$, including systematic uncertainties, as shown in Figure~\ref{fig:cs_charm}c. The branching fraction $\mathcal{B}(\psi(4160) \to K^-\bar{\Xi}\Lambda + \text{c.c.})$ is determined to be $(4.4\pm2.1)\times10^{-6}$. Similarly, upper limits on $\Gamma_{ee}\mathcal{B}$ are provided for other assumed resonances.
In the processes $e^+e^- \to pK^-\Lambda$, $\phi\Lambda\bar{\Lambda}$, and $\eta\Lambda\bar{\Lambda}$, clear enhancements can be seen in the invariant mass distributions of $p\Lambda$ and $\Lambda\bar{\Lambda}$~\cite{BESIII:2021fqx,BESIII:2022tvj,BESIII:2023kgz}. Based on fits with Breit–Wigner (BW) functions, the masses and widths of these near-threshold structures were determined, but they do not match any previously known resonances. Future studies with larger data samples and theoretical analyses involving partial wave analysis may help clarify the nature of these observed structures.\vspace{-12pt}

In addition, some studies have focused on the multi-body processes $e^+e^- \to 2(p\bar{p})$~\cite{BESIII:2020svk}, $e^+e^- \to pp\bar{p}\bar{n}\pi^- + \text{c.c.}$~\cite{BESIII:2022acw,BESIII:2024pue}, and $e^+e^- \to pK^-K^+\bar{\Xi}^+$~\cite{BESIII:2025lbj}, aiming to search for possible deuteron, hexaquark, or di-baryon candidates, as well as to explore charmonium(-like) states. Among these, the measurement results of the Born cross sections for the four-body processes are shown in Figure~\ref{fig:cs_34body} (right).
However, with the current statistics, no definitive conclusions can be drawn regarding the existence of actual resonances or structures in these analyses.

\section{Hyperon Nucleon Interaction}

\begin{figure}[H]

\includegraphics[width=0.8\textwidth]{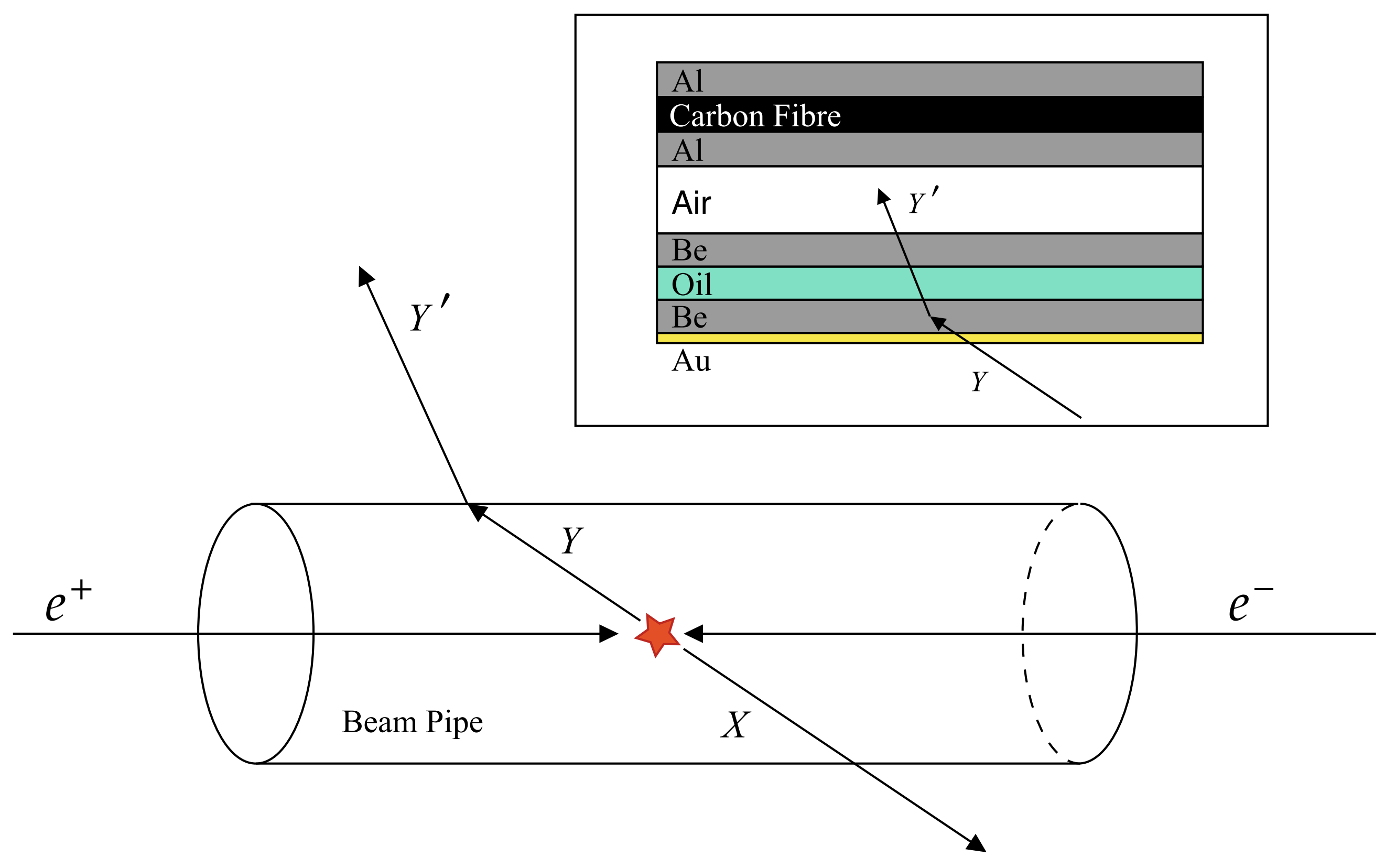}
\caption{\label{fig:pipe} Schematic of the hyperon--nucleus interactions at the $e^+e^-$ collider represented by BESIII~\cite{Dai:2024myk}. $Y$ and $Y'$ denote the hyperon before and after scattering, respectively, while $X$ represents the accompanying particles produced together with $Y$. }
\end{figure}

Depending on the charmonium decay modes producing hyperons, and using neutrons in the beam-pipe material as targets, BESIII has pioneered a series of hyperon–nucleon scattering studies.
The beam pipe, composed of gold ($^{197}$Au), beryllium ($^{9}$Be), and oil ($^{12}$C : $^1$H = 1 : 2.13), provides a natural and effective target for such studies~\cite{Dai:2024myk,Yuan:2021yks}.
The hyperon–nucleon scattering process in an electron–positron collider is illustrated in Figure~\ref{fig:pipe}~\cite{Dai:2024myk}. Hyperon pairs or final states containing hyperons are produced inside the beam pipe and travel along their momentum direction. Some of these hyperons reach the target before decaying and undergo elastic or inelastic scattering with the nuclei in the material.
However, due to the complexity of the materials in the beam pipe and the inner wall of the MDC, the BESIII experiment cannot directly extract hyperon–nucleon scattering cross sections as in fixed-target experiments. Therefore, $^{9}$Be, which is commonly used in fixed-target experiments and is also the primary material of the beam pipe, is chosen as the normalization reference, and the cross sections of all materials are normalized to that on~$^{9}$Be.

\begin{figure}[H]
 
\includegraphics[width=1.0\textwidth]{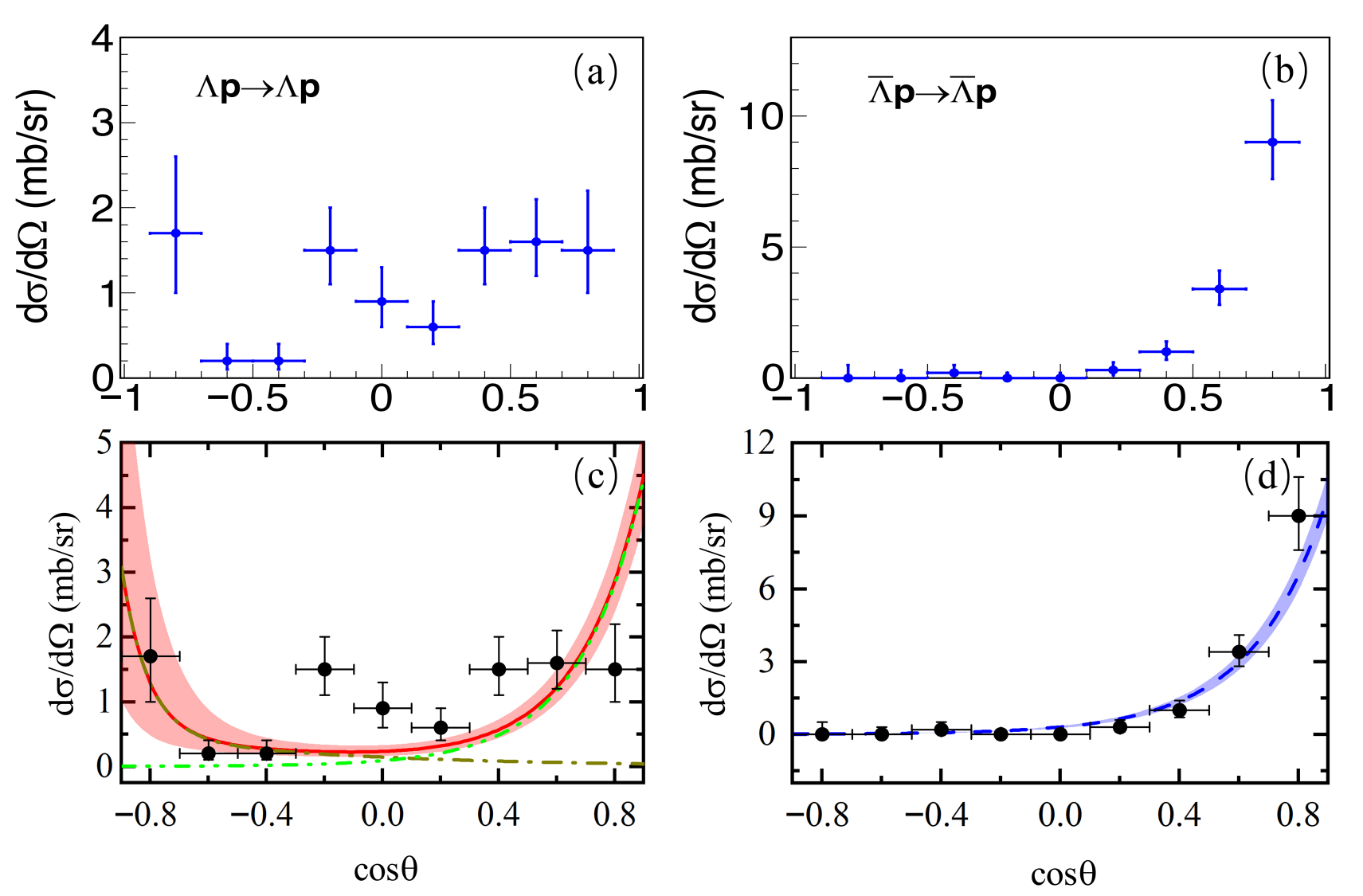}
\caption{Differential cross sections for the reactions $\Lambda + p \to \Lambda + p$ (\textbf{a},\textbf{c}) and $\bar{\Lambda} + p \to \bar{\Lambda} + p$~(\textbf{b},\textbf{d})~\cite{BESIII:2024geh,Wang:2024whi}. The dots with error bars represent the experimental data.
The solid red curve denotes the complete theoretical model, while the green dash–double-dotted and dark-yellow dash–dotted curves show the individual contributions from the $t$ and $u$ channels, respectively. The $t$-channel contribution is also displayed by the blue dashed curve.}\label{fig:LamP}
\end{figure}

The first study in this series observed the inelastic scattering process $\Xi^0 n \to \Xi^- p$ with a significance of 7.1$\sigma$, based on $\Xi^0$ hyperons produced in the decay $J/\psi \to \Xi^0 \bar{\Xi}^0$~\cite{BESIII:2023clq}. The reaction cross section of $\Xi^0$ with the $^9{\rm Be}$ target in the beam pipe was measured at a $\Xi^0$ momentum of 0.818 GeV/$c$. Assuming an effective number of three reaction neutrons in a $^9{\rm Be}$ nucleus, the cross section for the $\Xi^0 n \to \Xi^- p$ process was determined.
The inelastic scattering process $\Lambda + {}^{9}\mathrm{Be} \to \Sigma^+ + X$ was also investigated~\cite{BESIII:2023trh}. The cross section was measured within the $\Lambda$ momentum range of [1.057, 1.091] GeV/$c$. The\linebreak   $\Lambda + p \to \Sigma^+ + X$ cross section was further determined under the assumption that the signal originates from a surface reaction. The result is consistent with previous experimental measurements.
For the $\Lambda$ hyperon, an elastic scattering study was also performed for the processes $\Lambda + p \to \Lambda + p$ and $\bar{\Lambda} + p \to \bar{\Lambda} + p$, representing the first investigation of antihyperon–nucleon scattering~\cite{BESIII:2024geh}. Within the angular region $-0.9 \leq \cos\theta_{\Lambda/\bar{\Lambda}} \leq 0.9$ and at a $\Lambda$/$\bar{\Lambda}$ momentum of $1.074 \pm 0.017$ GeV/$c$, the cross sections for these two processes were measured. Here, $\theta_{\Lambda/\bar{\Lambda}}$ denotes the scattering angle.
However, the angular distributions of the cross sections for these two reactions exhibit distinctly different trends, as shown in Figure~\ref{fig:LamP}a,b. Theoretical studies attribute the observed difference to reaction dynamics~\cite{Wang:2024whi}: the $\Lambda p$ process involves contributions from both the $t$ and $u$ channels, while the $\bar{\Lambda}p$ process allows only the $t$-channel, leading to the distinct trends observed. Here, the theoretical predictions are in good agreement with the experimental data, as shown in Figure~\ref{fig:LamP}c,d.
A recent study has measured the inelastic scattering processes $\Sigma^+ n \to \Lambda p$ and $\Sigma^+ n \to \Sigma^0 p$~\cite{BESIII:2025bft}. At a $\Sigma^+$ momentum of $0.992 \pm 0.015$ GeV/$c$, the reaction cross sections with the $^9$Be nucleus were measured, and assuming an effective number of three reaction neutrons, the single-neutron cross sections were determined to be in good agreement with theoretical predictions.
The aforementioned measurements of the scattering cross section are summarized in Table~\ref{tab:scatter}.
\vspace{-9pt}

\begin{table}[H]
    \caption{The measured cross sections of hyperon–nucleon scattering~\cite{BESIII:2023clq,BESIII:2023trh,BESIII:2024geh,BESIII:2025bft}. The first uncertainty is statistical and the second is systematic.}
    
   \setlength{\tabcolsep}{18.3mm}{\begin{tabular}{l r}
    \toprule
        \textbf{Reaction}	                    &  \boldmath{$\sigma$} \textbf{(mb)}      \\
     \midrule
        $\Xi^0+^9{\rm Be} \to \Xi^-+p+^8{\rm Be}$           & $22.1\pm5.3\pm4.5$ \\
        $\Xi^0+n \to \Xi^-+p$                               & $7.4\pm1.8\pm1.5$  \\    
        $\Lambda+^{9}{\rm Be} \to \Sigma^++{\rm X}$         & $37.3\pm4.7\pm3.5$ \\
        $\Lambda+p \to \Sigma^++{\rm X}$                    & $19.3\pm2.4\pm1.8$ \\
        $\Lambda+p \to \Lambda+p$                           & $12.2\pm1.6\pm1.1$ \\
        $\bar\Lambda+p \to \bar\Lambda+p$                   & $17.5\pm2.1\pm1.6$ \\
        $\Sigma^++^{9}{\rm Be} \to \Lambda+p+^8{\rm Be}$    & $45.2\pm12.1\pm7.2$\\
        $\Sigma^++n \to \Lambda+p$                          & $15.1\pm4.0\pm2.4$ \\
        $\Sigma^++^{9}{\rm Be} \to \Sigma^0+p+^8{\rm Be}$   & $29.8\pm9.7\pm6.9$ \\
        $\Sigma^++n \to \Sigma^0+p$                         & $9.9\pm3.2\pm2.3$  \\
      \bottomrule
        \end{tabular}}
    \label{tab:scatter}
\end{table}

\section{Conclusions and Outlook}
Having been operating successfully since 2008, the BESIII experiment has collected the world's largest data samples in the $\tau$-charm physics region and has conducted a comprehensive program in hyperon physics. 
In studies of hyperon polarization, the angular distribution analysis method has been widely applied. Numerous polarization signals have been observed, $CP$ violation observables have reached the $10^{-3}$ level, studies of form factors and radiative decays have provided insights into the interaction mechanisms, and the measurement of the $\Lambda$ EDM has achieved unprecedented precision.
In the field of rare hyperon decay studies, baryon and lepton number violating decays as well as semileptonic decays have been extensively investigated, providing stringent experimental constraints and precision tests of the SM.
Extensive cross section measurements have been performed from the hyperon pair production threshold up to the charmonium energy region, including determinations of the relevant form factors for hyperon pair processes. Evidence for the decays $\psi(3770) \to \Lambda\bar\Lambda / \Xi^-\bar\Xi^+$ and $\psi(4160) \to K^-\bar\Xi^+\Lambda$ has also been reported. Moreover, an innovative experimental approach has been developed at BESIII to study hyperon–nucleon interactions, which has already been applied to various elastic and inelastic scattering processes.
These studies have greatly enriched our understanding of the internal structure and interactions of hyperons. Looking ahead, the BESIII Collaboration is updating its hardware and continuing data collection, with several ongoing exciting analyses; more fascinating results are expected in the near future.

\vspace{6pt} 
\authorcontributions{Conceptualization and methodology, R.Z. and X.W. All authors have read and agreed to the published version of the manuscript.}

\funding{This work was supported by 
the Fundamental Research Funds for the Central Universities Nos. lzujbky-2025-it06, 
lzujbky-2025-ytA05, lzujbky-2024-jdzx06;
the Natural Science Foundation of Gansu Province No. 22JR5RA389, No. 25JRRA799;
the ‘111 Center’ under Grant No. B20063; and
the National Natural Science Foundation of China under Contract No. 12247101.}

\dataavailability{No new data were created or analyzed in this study. Data sharing is not applicable to this article.} 

\conflictsofinterest{The authors declare no conflicts of interest.} 



\abbreviations{Abbreviations}{
The following abbreviations are used in this manuscript:
\\

\noindent 
\begin{tabular}{@{}ll}
$CP$  & Charge-conjugation and parity               \\
CKM   & Cabibbo–Kobayashi–Maskawa                   \\
BNV   & Baryon-number-violating                     \\
EoS   & Equation of state                           \\
c.m.  & Center-of-mass                              \\
MDC   & Multilayer drift chamber                    \\
TOF   & Time-of-flight system                       \\
EMC   & Electromagnetic calorimeter                 \\
c.c.  & Charge-conjugation                          \\
PHSP  & Phase space                                 \\
MC    & Monte Carlo                                 \\
VMD   & Vector meson dominance model                \\
PM    & Pole model                                  \\
NRCQM & Nonrelativistic constituent quark model     \\
BSU(3)& Broken SU(3) model                          \\
ChPT  & Chiral perturbation theory                  \\
EDM   & Electric dipole moment                      \\
C.L.  & Confidence level                            \\
DT    & Double-tag                                  \\
ST    & Single-tag                                  \\
LFU   & Lepton flavor universality                  \\

\end{tabular}
}

\begin{adjustwidth}{-\extralength}{0cm}

\reftitle{{References}}

\PublishersNote{}

\end{adjustwidth}
\end{document}